\documentclass[twocolumn,preprintnumbers,floatfix,prb,superscriptaddress,showpacs]{revtex4}
\usepackage{graphicx}
\usepackage{dcolumn} 
\usepackage{bm}
\usepackage{amsmath,amssymb}
\usepackage{epsfig}
\begin{document} 

\title{Dielectric Response of Electron-doped Ionic Superconductor Li$_x$ZrNCl}
  
\author{Antia S. Botana}
\affiliation{Department of Physics, University of California, Davis, CA 95616, USA}
\author{Warren E.~Pickett}
\affiliation{Department of Physics, University of California, Davis, CA 95616, USA}

\date{\today}
\begin{abstract}
When electron doped, the layered transition metal nitrides ${\cal T}$NCl
(${\cal T}$ = group IVB transition metal ion) become impressive superconductors with critical
temperature T$_c$= 15-26K. Here we take the most studied member, ZrNCl, as a
representative and calculate the dielectric response $\epsilon(\omega)$
versus frequency and concentration
of doped electronic carriers. The static dielectric constant $\epsilon_{\infty}$=5 is
reproduced extremely well. We establish that the differences between rigid band
modeling and virtual crystal treatment are small, and compare also with
actual lithium doping using supercells. We obtain the variations upon changing the doping level of the reflectivity and energy loss function as well,
many of which are found not to be correlated with the observed (non)variation of T$_c(x)$. The main
plasmon peaks appear where the electron gas model suggests, in the range 1.2-2.0 eV for
$x$ varying from 0.16 to 0.50. 
\end{abstract}


\maketitle

\section{Introduction and Background}

Superconductivity in doped quasi-two-dimensional compounds or at surfaces has been a focus of interest of the community since the work of Ginzburg,\cite{ginzburg_1, ginzburg_book} and has gained fevered activity after the discovery of the strong dependence of the critical temperature (T$_c$) upon intercalation of the transition metal dichalcogenides and the discovery of high T$_c$ in 2D cuprates. The way in which doping of 2D layers modifies the electronic behavior of a compound giving rise to superconductivity greatly depends on the specific system. In copper oxides there is a magnetic insulator to strange metal transition that completely modifies the electronic structure. However, in bismuthates and in electron-doped nitridochloride $A_x{\cal T}$NCl superconductors (where ${\cal T}$ is a group IVB transition metal Ti, Zr, Hf) there is only Pauli  paramagnetic behavior. The cause of superconductivity in this latter class of materials remains unknown, 
with neither spin nor conventional phonons being responsible for their superconducting behavior, as we discuss below. \cite{pickett_1}

\begin{figure}
\begin{center}
\includegraphics[width=4.6cm,draft=false]{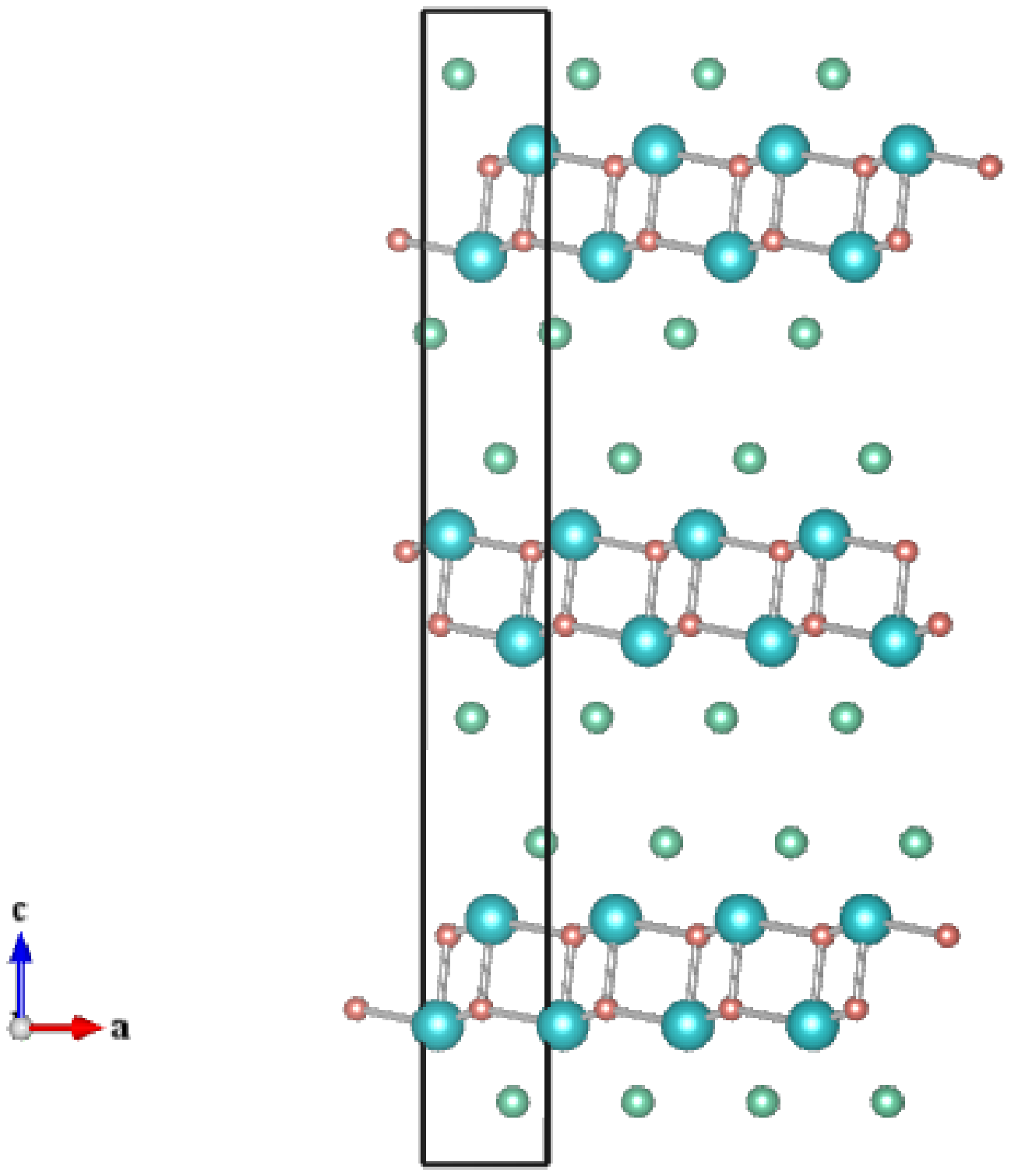}
\includegraphics[width=2.95cm,draft=false]{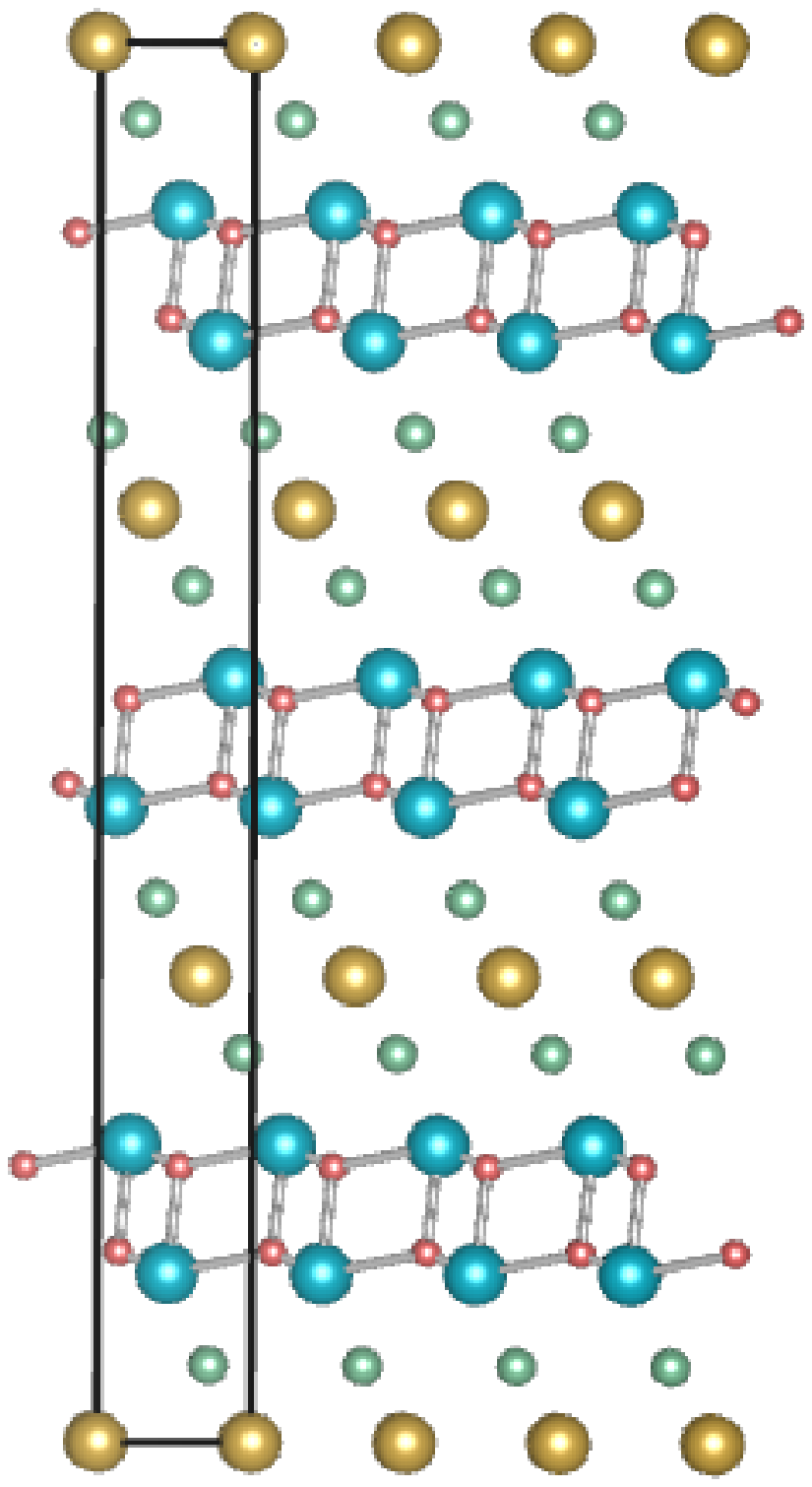}

\caption{Crystal structure of ZrNCl (left) and the doped compound Li$_{0.5}$ZrNCl (right). Zr atoms are blue, N red, Cl green and Li yellow.}\label{zrncl_struct}
\end{center}
\end{figure}

The superconductivity in transition metal nitridochlorides (with maximum T$_c$ = 17, 15, 26 K for   Ti, Zr, Hf respectively)  emerges when the insulating parent compounds are electron-doped.\cite{yamanaka_nat_hfncl, yamanka_tincl, yamanka_zrncl} To clarify the mechanism of superconductivity an understanding of the electronic behavior is crucial because only then can the key issue of electron pairing be addressed. Electron doping in the metal nitride double honeycomb planes, as shown in Fig. \ref{zrncl_struct}, is achieved by  intercalation of alkali ions into the van der Waals gap between Cl layers. Large organic molecules may also be included.\cite{interlayer_coupling_2,tincl_res,interlayer_coupling_1} However, the origin of the high T$_c$ remains unknown, as the characteristics of this class of superconductors are unique. There is a low DOS at the Fermi level and a weak electron phonon coupling has been suggested both experimentally \cite{iwasa_heat_2, iwasa_tc_enhancement_doping} and theoretically. \cite{bohnen, weht}

There are several unusual characteristics that make the $A_x{\cal T}$NCl class of superconductors unique.
First, and an aspect that is not usually emphasized, they are extremely bad conductors, with reported residual
resistivity (T$\rightarrow$0) of $10^3-10^5$ $\mu \Omega$ cm.\cite{iwasa_prb, interlayer_coupling_2, interlayer_coupling_1, yamanaka_nat_hfncl, tincl_res}
In some cases the temperature dependence crosses from metallic to
semiconducting and weak localization behavior arises just above T$_c$. Second, except near the
critical concentration $x_{cr}$ where the insulator to metal
transition occurs,\cite{iwasa_tc_enhancement_doping} T$_c$ is almost independent of the doping level $x$ whereas the Fermi level DOS undergoes a change as doping is increased. 
Specifically, there is no superconducting dome, and
the constancy of T$_c$ continues well into the region where
the density of states at the calculated Fermi level $N(E_F)$ increases sharply, as the
${\cal T}$ ion $d$ states other than the in-plane $d_{xy},
d_{x^2-y^2}$ orbitals begin to be filled. \cite{dft3} Thirdly, there is one property
that correlates strongly with T$_c$: the separation
of the ${\cal T}N$ bilayers along the $z$ direction. 
For the Hf and Zr counterparts, from the minimum separation, at small Li intercalation, T$_c$ increases
sharply by $\approx$30\% before leveling off and remaining nearly constant up to the largest separations that have
been achieved ($\sim$ 20 \AA), by introducing large molecules along with the
dopant metal atoms.\cite{interlayer_coupling_2,interlayer_coupling_1} Thus the superconductivity in these systems is truly 2D. 

The very ``bad conductor'' aspect can be clarified by trying to apply
conventional Fermi liquid metallic formalism. Interpreting the residual resistivity 
$\rho_{\circ} = 4\pi/\Omega_p^2 \tau$ $\sim$ 10$^4$ $\mu$$\Omega$ cm
gives, with in-plane $\hbar\Omega_p \approx$ 3 eV, $\hbar/\tau \sim$
10 eV, corresponding to a totally washed out band structure. 
Putting $\tau$ together with the Fermi velocity $v_F \approx$ 10$^8$ cm/s
leads to a mean free path ($l$= 2$\pi$/$\Delta$$k$) that implies the Brillouin zone (BZ) would be completely wiped out. This analysis indicates the transport is not metallic, and an alternative picture is
needed to understand electronic behavior.

Takano \textit{et al.}, on the other hand, use the measured superconducting coherence length $\xi$, the BCS relation for $\xi_0$, and the phenomenological relation $\xi^{-1}$= $\xi_0^{-1}$+ $l$$^{-1}$ to infer values of $l$$\sim$ 10 nm for their samples. \cite{iwasa_prb} In any case, the lack of consistency in these interpretations indicates problems with Fermi liquid theory for transport.
This conclusion gains support from the recent report from $^{91}$Zr and $^{15}$N NMR studies\cite{nmr}, which finds pseudogap
character typical of non-Fermi liquid phases in the 0.06$\leq$$x$$\leq$0.10 regime. This is the regime where enhanced susceptibility is reported;\cite{enhancement_pairing} curiously, the NMR
data show no evidence of enhanced spin fluctuations on the verge of the superconductor to insulator transition, causing difficulties for applying a model of pairing by this mechanism. Kuroki proposed that the high T$_c$ in doped Hf/ZrNCl is a characteristic feature of the spin fluctuation-mediated superconductivity in a honeycomb lattice.\cite{kuroki}

Superconductivity emerges in this system from a very badly conducting state. 
Jaramillo \textit{et al.}\cite{jaramillo} have proposed that a strong 
interaction that can shift spectral weight over a large energy scale is 
necessary for bad metal conductivity. In the case of rare-earth nickelates, 
this strong interaction happens between lattice distortions and Ni-O 
covalence, but its origin may vary depending on the system (and nickelates 
do not superconduct). The question of whether this explanation may apply to 
the bad conductivity in $A_x{\cal T}$NCl superconductors and if so, 
what the driving interaction might be, is still open.

There is no local moment behavior, so the electronic state and
the pairing mechanism are very different from the high
temperature superconducting cuprates and Fe-based materials.
Based on the standard Landau Fermi liquid (metallic) viewpoint, the
electron-phonon coupling has been calculated to be clearly too weak to
account for the observed values of T$_c$. \cite{bohnen, weht} 
From density functional theory (DFT) based calculations for Li-doped ZrNCl, 
$T_c$ of only 5 K is obtained, using a typical value of the Coulomb 
pseudopotential for a normal metal $\mu^*$= 0.1.\cite{bohnen} 
A suggestion was made by Yin {\it et al.}\cite{kotliar_hybrid} 
that a different exchange-correlation functional, the hybrid (part
unscreened exchange) functional, is more appropriate for this class
of doped insulators. They found that the additional exchange potential increases the
shifts of certain bands due to selected frozen phonon modes, giving them
an enhanced electron-phonon coupling. However, a full calculation was
not carried out. In any case, whichever functional is chosen,
it seems apparent from the observed resistivity that Li$_x$ZrNCl 
has no transport Fermi surface at all,
or very few mobile carriers, or both. 

The N isotope shift of T$_c$
seems to be non-zero but is so small as to be very discouraging for 
any lattice mechanism relying on N displacements. \cite{n_isotope_shift_hf,n_isotope_shift_zr}
The density functional theory for superconductors (SCDFT)\cite{scdft} 
has been shown to reproduce the observed T$_c$ of a selection of 
phonon-mediated superconductors.\cite{scdft_mgb2, scdft_graphite, scdft_li_al} 
This approach differs largely in the DFT-based treatment of the Coulomb
interaction. When SCDFT was applied to ${\cal T}$NCl superconductors,\cite{dft3} a T$_c$ of 
only 4.3 K and 10.5 K was obtained for the Zr and Hf counterparts, respectively. 
In addition, T$_c$ increases with increasing doping level in clear contrast to 
the experimental results. In Li$_x$ZrNCl, an enhancement of T$_c$ with a 
reduction in the carrier density in the low doping regime towards the metal-insulator transition
has been observed.\cite{iwasa_tc_enhancement_doping}

Several works have tried to consider the origin of pairing in strongly 
layered superconductors beyond the standard Migdal-Eliashberg ansatz. 
Bill \textit{et al.}\cite{dynamical_screening, bill} have considered that the dynamical screening of the Coulomb 
interaction is essentially different in layered structures, and built
a model that provides for an additional contribution to the pairing 
from dynamical electronic screening and low energy plasmons, which leads 
to a drastic enhancement of T$_c$ in ${\cal T}$NCl 
superconductors. We also mention the work of
Pashitskii and Pentegov\cite{pashitskii} that has included classes
of vertex corrections within the picture of a plasmon mechanism of pairing.

To elucidate the electronic behavior and help to understand the origin of the unique properties of this class, we analyze the dependence of the dielectric response on frequency and concentration for Li$_x$ZrNCl calculated within the random phase approximation from the Kohn Sham band structure. This treatment is consistent with the Landau Fermi liquid picture which as we have noted is quite suspect for transport but may yet apply for higher energy excitations.

The paper is organized as follows. After describing the computational methods and crystal structure in Sections \ref{structure} and \ref{methods}, we revisit the electronic structure of the pristine compound and Li$_x$ZrNCl ($x=$ 0.16, 0.25 and 0.50), analyzed using supercells to reproduce the desired doping levels (Section \ref{es}). The obtained electronic structure of the doped compounds is then contrasted to that obtained using the virtual crystal approximation (VCA). In Section \ref{optic}, the dielectric response versus frequency and concentration is discussed. Differences in the dielectric behavior do not correlate with doping in the observed variation of  T$_c(x)$ suggesting that pairing based on the electronic overscreening is not a viable picture.

\section{Structure}\label{structure}

Although the compound $\beta$-ZrNCl has been known since the 60`s, its crystal structure was for some time under debate.\cite{structure_diff, structure_zrncl_1, structure_zrncl_2} We have used the structures determined by x-ray diffraction on single crystals\cite{structure_zrncl_2} that confirmed the results obtained by others using neutron powder diffraction.\cite{shamoto_structure}

The structure of the insulating parent compound $\beta$-ZrNCl  is shown in the left panel of
Fig. \ref{zrncl_struct}. The central structural units are double honeycomb layers of ZrN sandwiched between two Cl layers leading to a neutral ZrNCl unit. Adjacent ZrN layers are 
rotated such that a short Zr-N interlayer bond exists. Each Zr atom is 
bonded to four neighboring N, three belonging to the same ZrN layer and 
one to the adjacent layer. The bonding between the ZrNCl units is of 
a weak van der Waals type, allowing intercalation by alkali ions and also
by large organic molecules. Neighboring Zr$_2$N$_3$Cl$_2$ slabs are shifted 
relative to each other, resulting in a rhombohedral space group $R{\bar 3}M$.

Upon Li doping, the dopants occupy a high symmetry site ($3a$) within the van der Waals gap between two Cl layers (see right panel of Fig. \ref{zrncl_struct}).  As there is one such $3a$ site per bilayer, full occupancy corresponds to a doping level of $x=$0.5. The space group is not changed but the shift between neighboring bilayers is altered in such a manner that the stacking sequence is changed from ABC to ACB upon Li intercalation (see Fig. \ref{zrncl_struct}).

\section{Computational Methods}\label{methods}

T$_c$ does not change for the range $x=$ 0.16 to 0.50 so this is the doping range we will focus on, performing calculations for $x=$ 0.16, 0.25 and 0.5.\cite{enhancement_pairing} The different doping levels have been achieved in two different ways: $(i)$ constructing supercells from the structural data for Li$_{0.21}$ZrNCl (full occupancy of the $3a$ site) introducing Li vacancies to reach the desired doping level (for $x=$ 0.16, $3\times 2\times1$, for x= 0.25, $2\times 2\times1$), $(ii)$ using virtual crystal approximation (VCA) to avoid the use of supercells. In this case an artificial doping is achieved changing the electron count to the desired level to account for a certain doping. A doping level of 0.16 would correspond to 1/3 occupation of the Li site and 0.25 to 1/2. Two models can be used: without Li, but with 1/2(1/3) electrons per unit cell added to simulate the Li doping of 0.25(0.16) starting from the structure of ZrNCl or with one Li per unit cell (full occupancy) but removing 2/3(1/2) electrons to reach the same doping levels of 0.16(0.25). \cite{bohnen}

The electronic structure calculations were performed with the WIEN2k code,\cite{wien2k,wien} based on density functional theory\cite{dft,dft_2} (DFT) utilizing the augmented plane wave plus local orbitals method (APW+lo).\cite{sjo} All structures were fully relaxed using the generalized gradient approximation (GGA) PBE scheme\cite{gga} and the lattice parameters were optimized within the same scheme. The lattice parameters used in the calculations for the doped compound were $a=$ 3.60 \AA ~and $c=$ 27.83 \AA; for the undoped compound, $a=$ 3.59 \AA ~and $c=$ 27.67 \AA.

The calculations were well converged
with respect to the k-mesh and R$_{mt}$K$_{max}$.  R$_{mt}$K$_{max}$= 7.0
was chosen for all the calculations. Selected muffin tin radii were
the following: 2.07 a.u. for Li, 2.42 for Cl, 2.00 for Zr and 1.72 a.u. for N.

The optical properties were obtained using the optic code implemented within WIEN2k. The theoretical background is described in detail in Ref. \onlinecite{claudia}. The calculation of optical properties requires a dense mesh of eigenvalues and the corresponding eigenvectors (up to 10000 k points were used for Li$_x$ZrNCl). 

The program optic generates the symmetrized squared momentum matrix elements
between all band combinations for each $k$-point,  then it carries out the BZ integration.
The interband and the intraband contributions to the imaginary part of
the dielectric tensor are discussed separately in Section \ref{optic}. 
The Kramers-Kronig transform real component can be computed allowing the evaluation of the optical conductivity, loss function, and reflectivity.\cite{claudia} 

\section{Electronic structure, with ZrNCl as the prototype}\label{es}

\subsection{Undoped ZrNCl}

\begin{figure}

\includegraphics[width=4.05cm,draft=false]{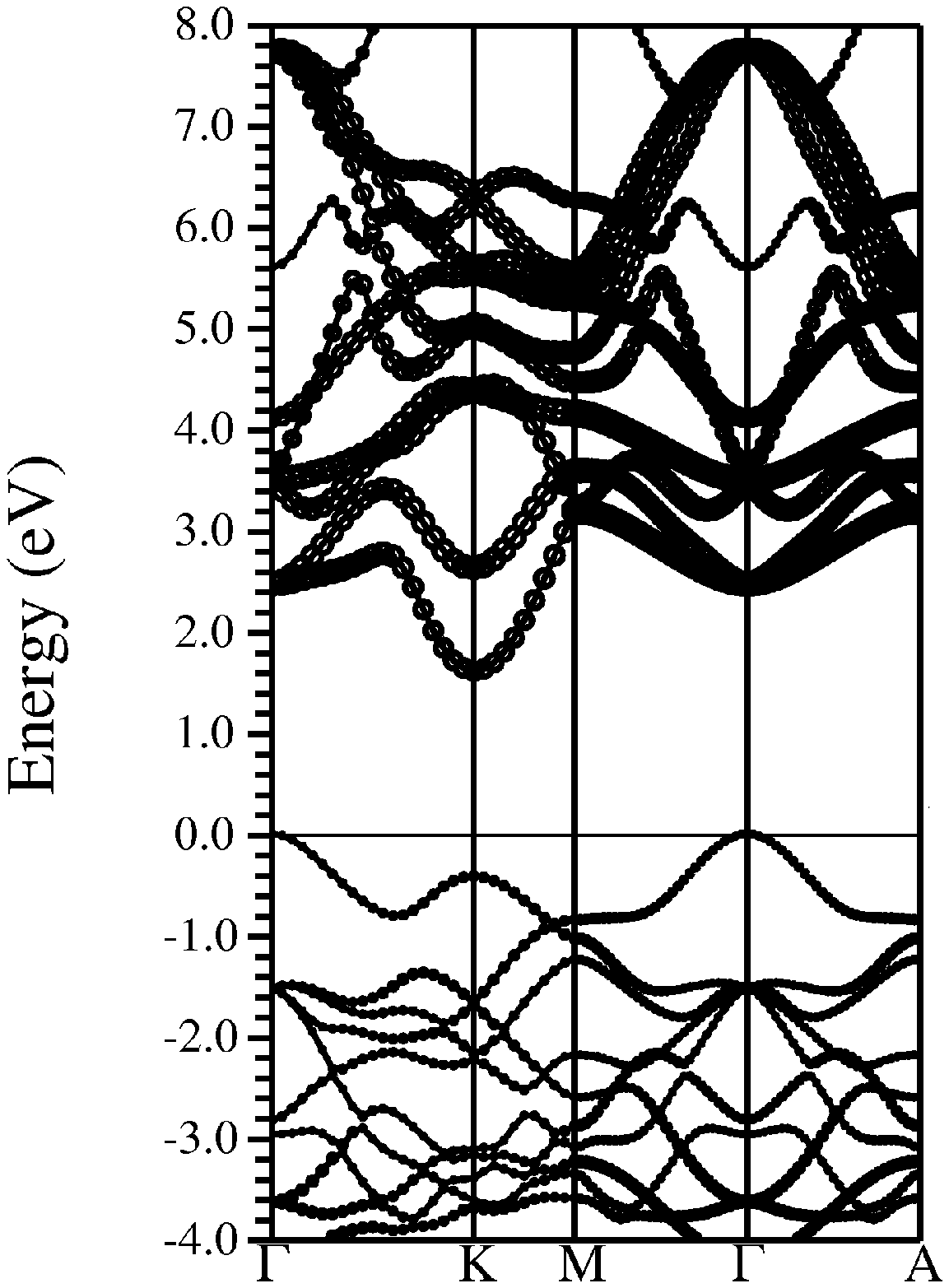}
\includegraphics[width=3.58cm,draft=false]{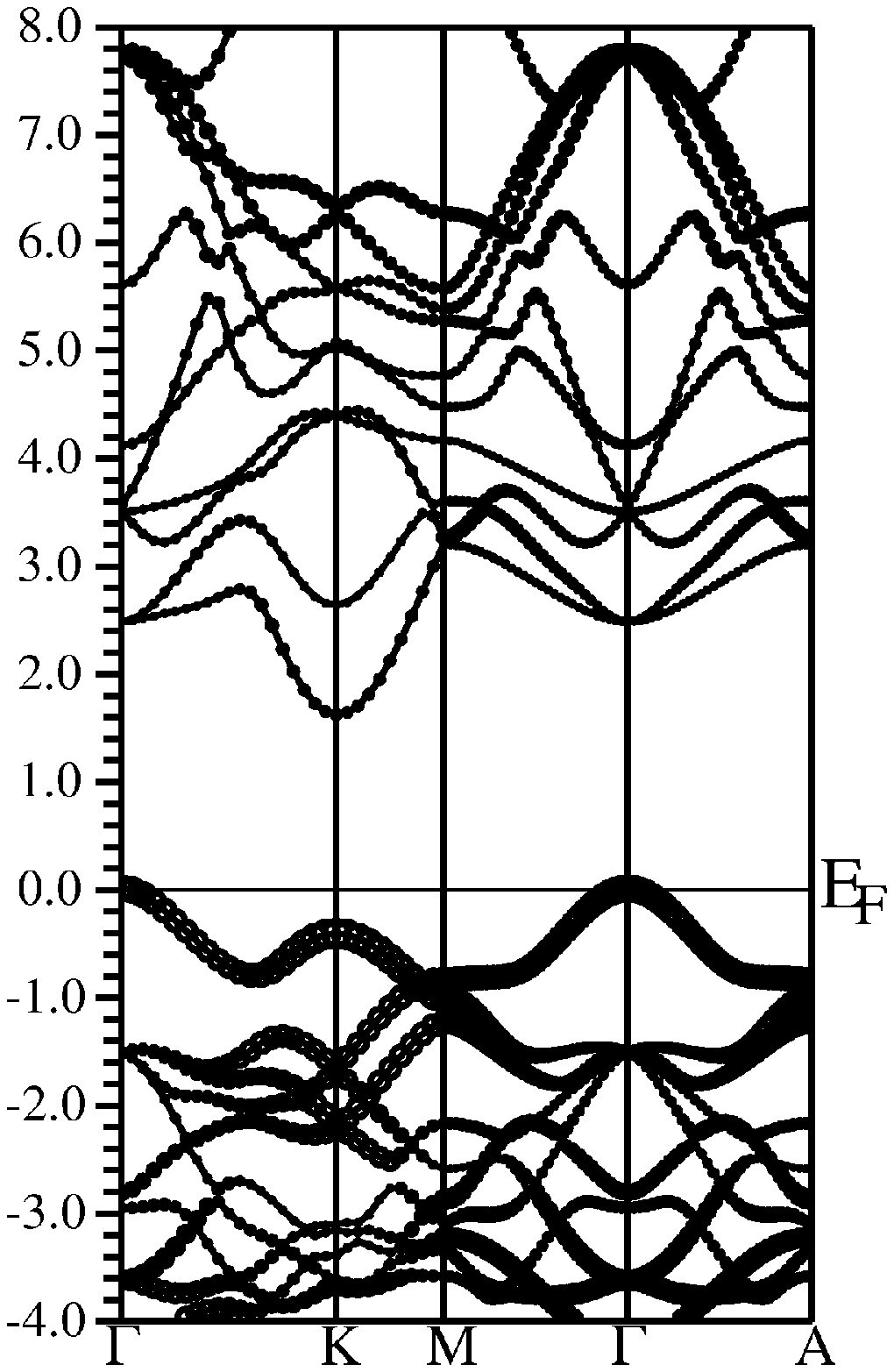}
\includegraphics[width=3cm,draft=false]{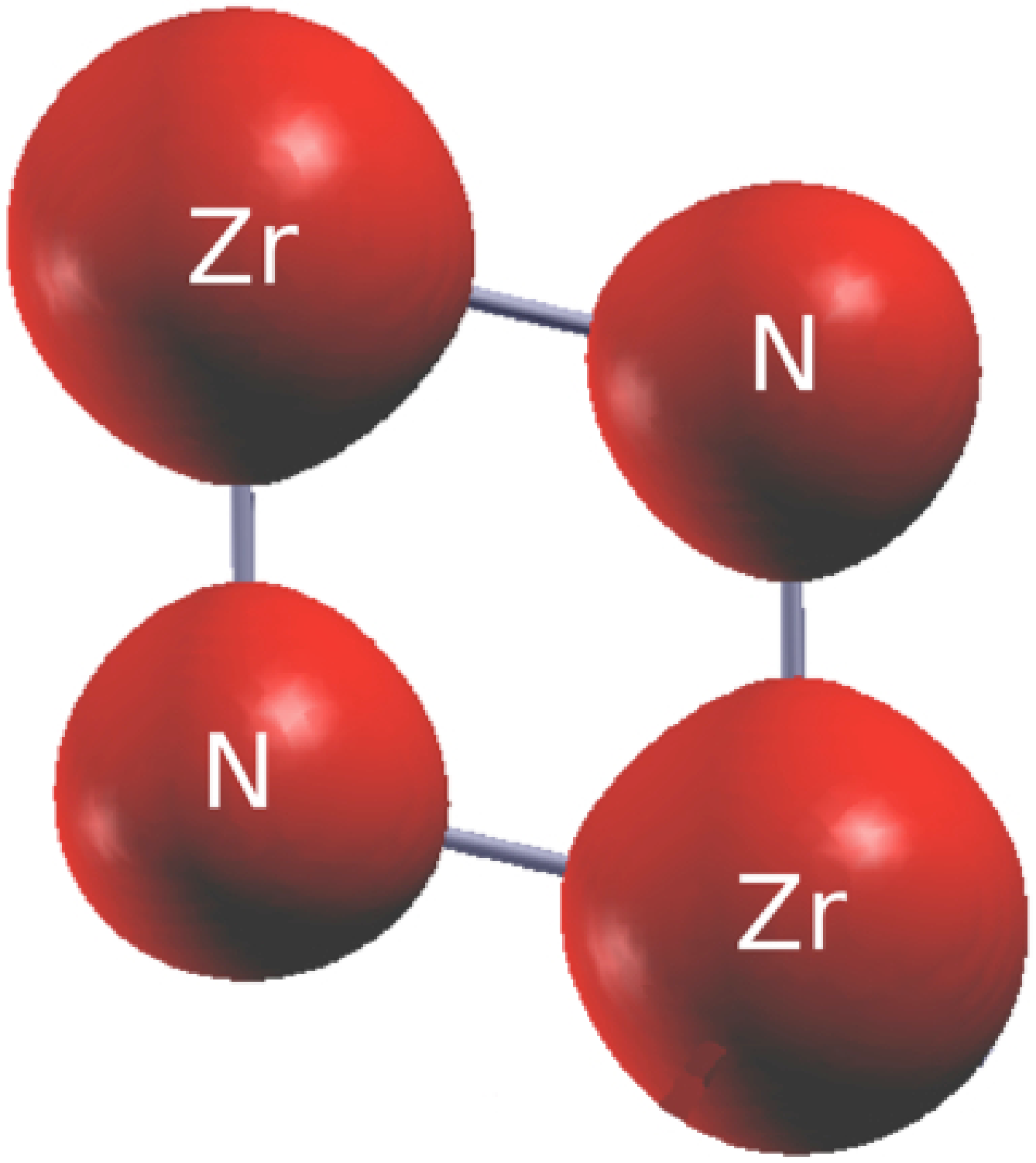}

\caption{Top panels. Band structure along high symmetry direction with band character emphasized, 
for ZrNCl. Left Zr $3d$, right N $2p$ character, is highlighted. The size of the symbols is 
proportional to the Zr or N character of the corresponding eigenfunction. Bottom panel: 
three dimensional isocontour plot of the charge density for the ZrN bilayer obtained using the 
Xcrysden package.\cite{xcrysden} The nearly spherical density obtained for both Zr and N atoms 
reflects that substantial Zr $3d$ character is mixed into the occupied N $2p$ bands, somewhat
reducing the ionic character from nominal values. }\label{bs_zrncl}
\end{figure}

The electronic structure of ZrNCl, with filled N $2p$ and Cl $2p$ bands and 
empty Zr $3d$ bands, is well characterized by the formal ionic description 
Zr$^{4+}$N$^{3-}$Cl$^{1-}$. Zr-N covalency is however substantial, so ZrNCl 
is quite different from a purely ionic insulator. In Fig. \ref{bs_zrncl} the 
band structure with band character plot for Zr and N atoms is shown. 
The band gap value for the parent compound determined experimentally is 2.5 eV,\cite{exp_bw}  the one derived within GGA is 1.7 eV. The conduction
band is dominated by empty Zr-$d$ states above the gap.
The $k_z$ dispersion along $\Gamma-A$ in
the undoped compound, which is comparable to in-plane dispersion, is due to 
the small interlayer distance. This dispersion will disappear when Li is
placed between the (ZrNCl)$_2$ slabs, thereby separating the layers and
practically eliminating overlap between them.

The Cl and N states are mixed
in the valence band region, though at the top of the valence band N states dominate.
The lowest lying conduction bands (at $K$) are formed precisely from the Zr $d_{xy}$, $d_{x^2-y^2}$
orbitals, and the bands just below the gap are N $p_x, p_y$ states.\cite{seshadri} With 3-fold in-plane coordination,
N $p_x, p_y$ orbitals should be thought analogously to the $sp^2$ orbitals in graphene.
Likewise, the Zr $d$ orbitals should be pictured in terms of 3-fold
symmetry adapted orbitals, for both pairs ($d_{x^2-y^2}$,$d_{xy}$) and $(d_{xz}, d_{yz}$). 
However, isovalent TiNCl with T$_c$=17 K has an orthorhombic 
structure,\cite{yin} so it seems
clear that the honeycomb structure of ZrNCl and HfNCl does not play a specific role
in the occurrence of high temperature superconductivity in this materials class.

The electron carriers go into states largely on Zr, with lobes of maximum charge pointing
{\it between} neighboring N ions, as the crystal field picture would suggest. The density
on the N ions, according to the same picture is oriented directly at the positive Zr ions. The
Zr-N hybridization is best characterized as antibonding. \cite{weht}
The $d$$_{xz}$,$d$$_{yz}$ and $d$$_{z^2}$ orbitals lie successively higher in energy.

\subsection{Li$_x$ZrNCl supercells}

\begin{figure*}
\begin{center}
\includegraphics[width=5cm,draft=false]{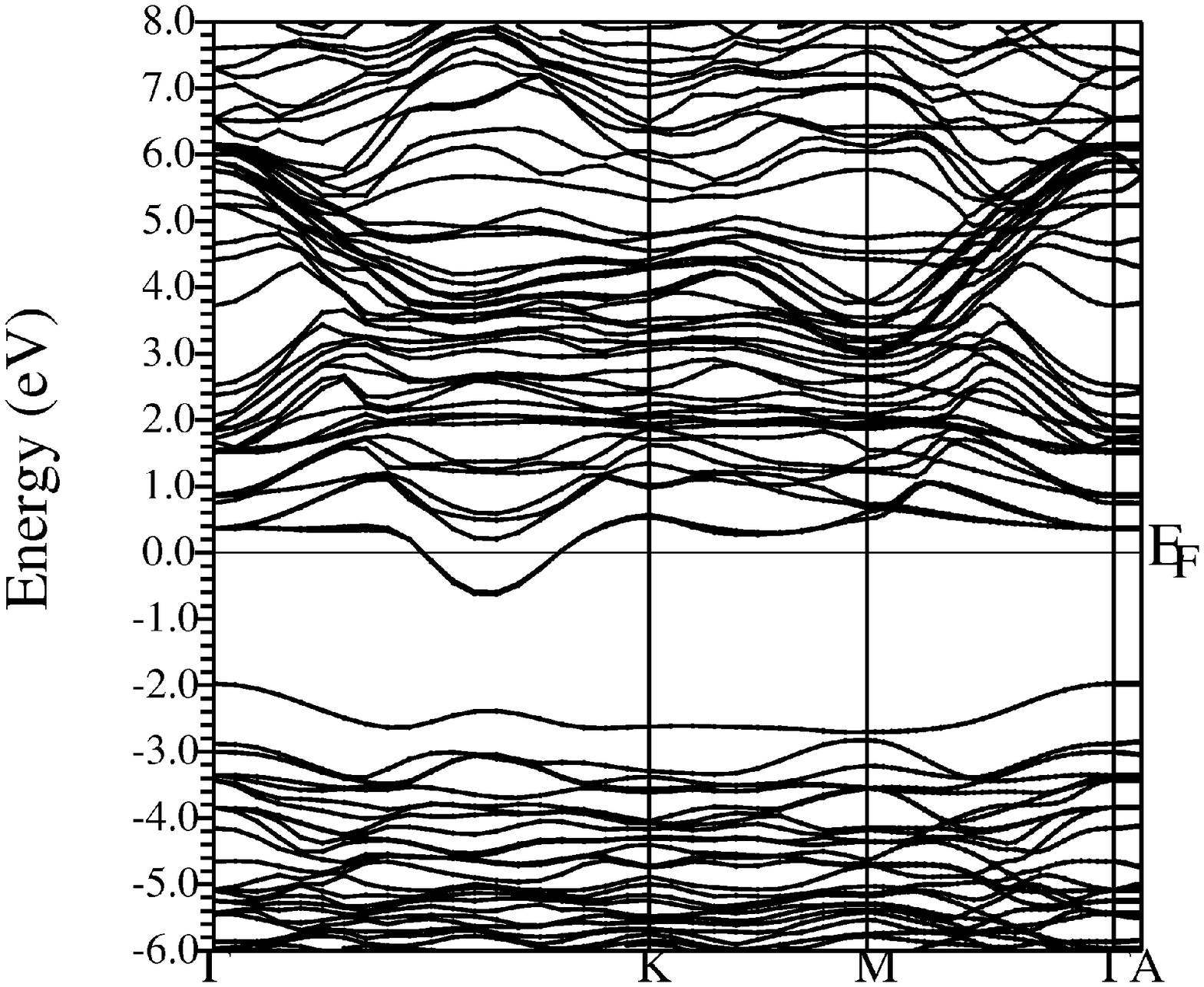}
\includegraphics[width=5cm,draft=false]{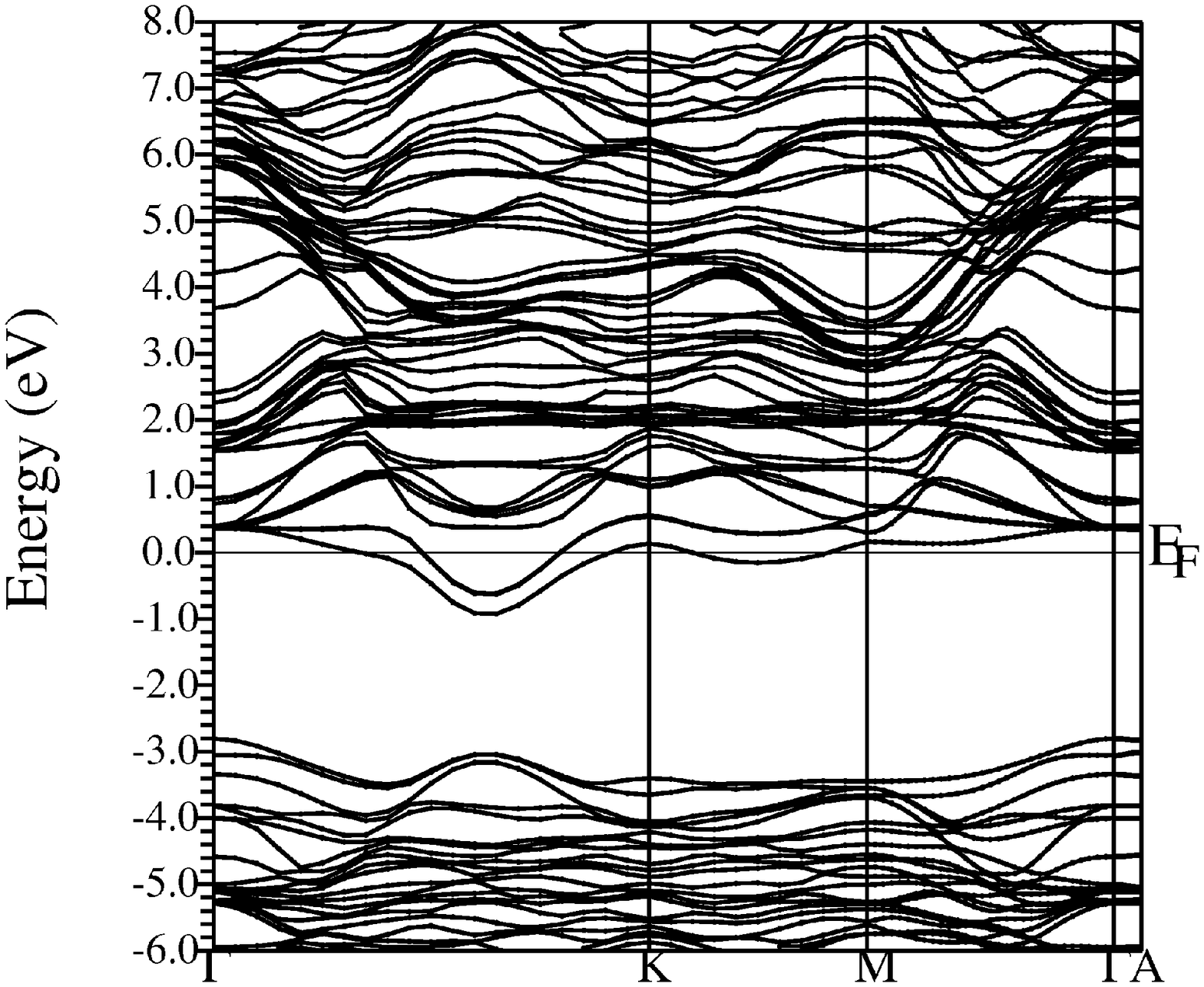}
\includegraphics[width=5cm,draft=false]{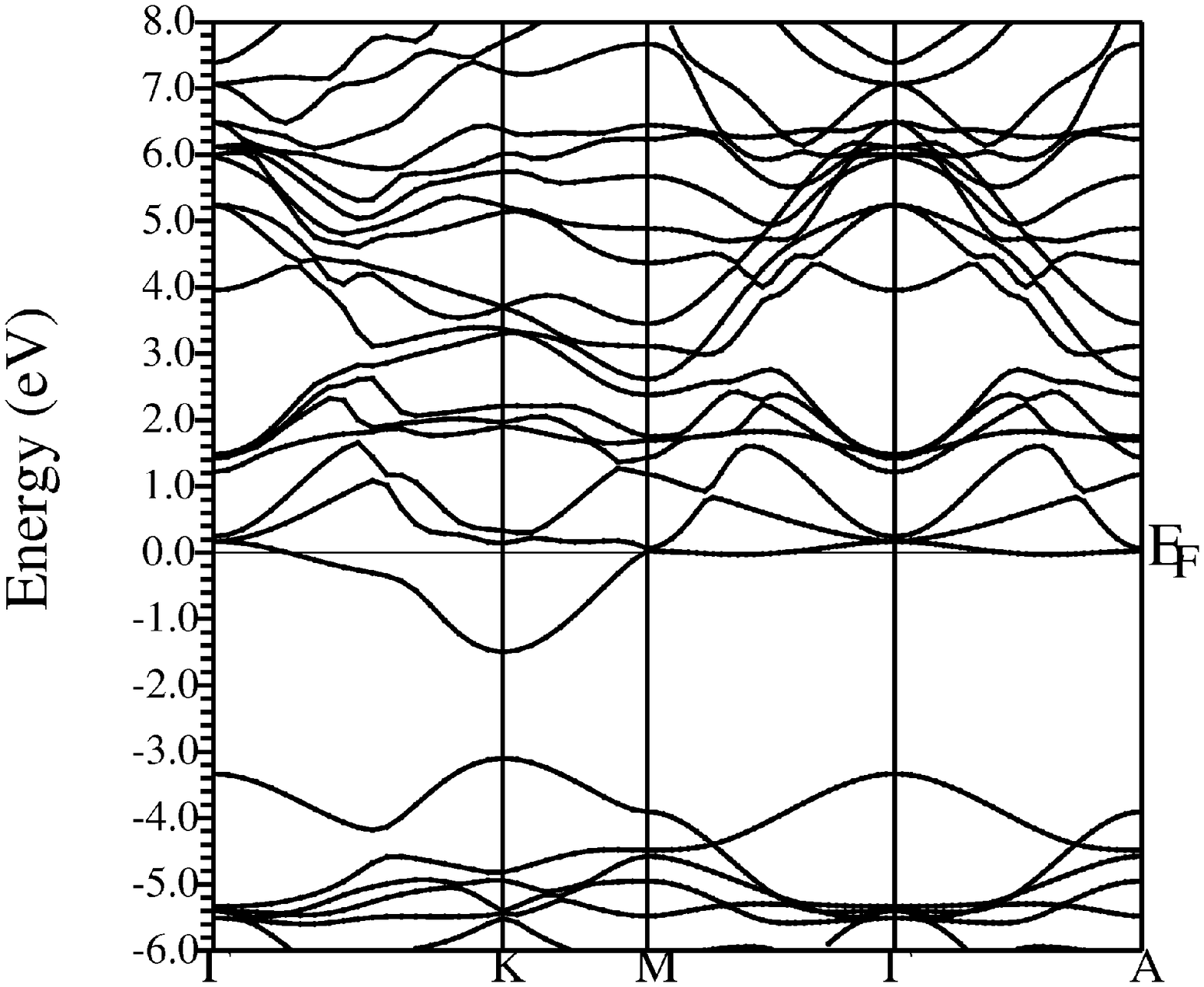}
\caption{Band structure of Li$_x$ZrNCl. From left to right $x$= 0.16, 0.25 (supercells) and 0.50 (not supercell).}\label{lix_bs_1}
\end{center}
\end{figure*}

The band structures for Li$_x$ZrNCl $x$= 0.16, 0.25 (using supercells) and $x$= 0.50 
are shown in Fig. \ref{lix_bs_1}. Neither the structural changes nor the occupation of 
the Li site seem to alter significantly the shape of the lowest conduction band except 
to greatly decrease $k_z$ dispersion due to separation of the constituent layers. 
The differences that are visible can be ascribed to supercell effects arising from the (artificial) periodicity.

\begin{figure}
\begin{center}
\includegraphics[width=0.97\columnwidth,draft=false]{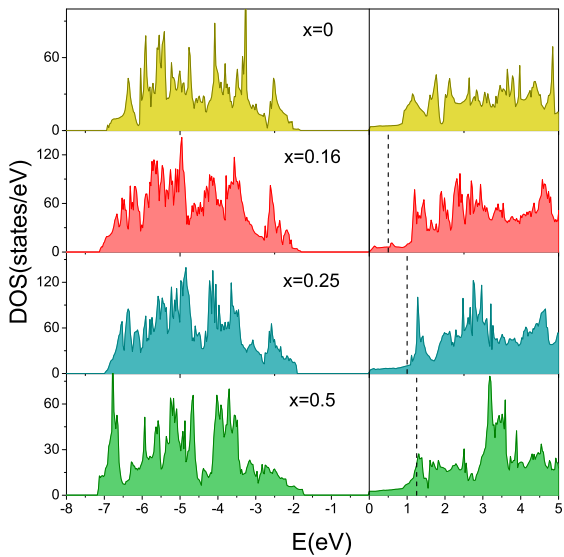}
\caption{Total DOS for ZrNCl and Li$_x$ZrNCl with increasing $x$ (from top to bottom). The zero energy is set at the bottom of the conduction band. The Fermi level is marked with a dashed line.  Note that the DOS at the top of the valence band changes, i.e. it is not rigid band like.}\label{dos_zrncl}
\end{center}
\end{figure}

The densities of states (DOS) for the various doping levels are shown in Fig. \ref{dos_zrncl} with the zero of energy set at the bottom of the conduction band. The lowest band is characterized by a small density of states at the Fermi level and a small effective mass as shown in Fig. \ref{dos_zrncl} ($N(E_F)$= 0.34 states/eV f.u  corresponds to an effective mass $m$$^*$=0.66 $m$).\cite{bohnen} The Fermi level (dashed lines) for the two lowest doping levels ($x$= 0.16 and 0.25) lies within this band with low DOS. For $x=$ 0.5, states other than the lowest in-plane $d$$_{x^2-y^2}$+$d$$_{xy}$ begin to be filled with correspondingly higher N(E$_F$). The valence bandwidths reported experimentally are 6.1 for the parent compound and 7.0 for Li$_{0.25}$ZrNCl,\cite{exp_bw} underestimated (which is not uncommon) by calculations performed using GGA. The DOS gives a clearer view of the fact that, upon electron doping, except for a shift in the chemical potential, the change in position of E$_F$ for higher doping levels, and some changes in the first peak in the DOS appearing at about 1 eV above the bottom of the conduction band, it remains unchanged. In the valence band region, bigger changes take place, the DOS at the top of the valence band changes not being only a rigid-band like shift. This difference should be irrelevant to superconductivity though.

The specific heat ($C_v$ ) jump at T$_c$ is a measure of N(0) (and the density of pairs that are formed). The observed extremely bad metallic character suggests that only a very small density of pairs may be formed at T$_c$. In BCS or Eliashberg-type materials, $\Delta C_v$ = 2-3 $\gamma T_c$, where $\gamma$ is the specific heat $\gamma$ proportional
to N(0) $\gamma$= 2/3(1+$\lambda$)$\pi^2k_B^2N(0)$. From the experimentally determined $\gamma$= 1.1 mJ/molK$^2$ using $N(0)= 0.19-0.26$ states/(eV spin f.u) a $\lambda \leq 0.22$ has been obtained by Iwasa \textit{et al.}\cite{iwasa_heat_1, iwasa_heat_2} This value belongs to the weak coupling regime and will never produce any finite value of T$_c$ once the Coulomb pseudopotential $\mu^*$= 0.10-0.15 is included. 



\subsection{Li$_x$ZrNCl within VCA}

Given the impressive superconductivity in this system and the uncertain nature of the pairing mechanism,
it is important to quantify the effect on the electronic structure of including Li explicitly,
 and on the method of doing so.
This aspect of charge rearrangement due to doping was found to be surprisingly 
large\cite{Ylvisaker} in (Al,Mg)CuO$_2$, another layered transition metal compound.
We have evaluated the electronic structure of Li$_x$ZrNCl within VCA 
using two alternative strategies: removing electrons from the stoichiometric structure of Li$_{0.5}$ZrNCl (one f.u. is Li(ZrNCl)$_2$) and adding electrons from ZrNCl. \cite{bohnen}
We find that actual occupation of the Li site does not alter the band structure
in any significant way, 
with carrier doping occurring in a rigid band fashion as has been assumed in all
previous works both theoretical (band structure calculations)\cite{bohnen} and experimental (optical reflectivity measurements). \cite{ iwasa_optic} For the lowest doping levels studied electrons will fill only the same single band with minimum at $K$ with Zr-($d_{x^2-y^2}$,$d_{xy}$) character using both VCA methods. For x= 0.5, some out of plane $d$-states start being occupied as shown in the previous section. In Na-doped ZrNCl and HfNCl, photoemission spectroscopy studies\cite{yok_1, yok_2} have shown non-rigid like behavior against theoretical predictions.\cite{weht}  The behavior appears as spectral weight shifts, whereas peak centroids remain rigid-band like.

The similarity of the electronic structures in the vicinity of the Fermi energy for empty and full occupation of the Li shows that, as far as the electronic properties are concerned, the Li atoms simply act as donators of electrons to the ZrN bilayers, which can justify a simpler doping study using VCA instead of the more computationally demanding use of supercells.

\subsection{Electron density plots for Li$_x$ZrNCl}

\begin{figure}
\begin{center}

\includegraphics[width=3.8cm,draft=false]{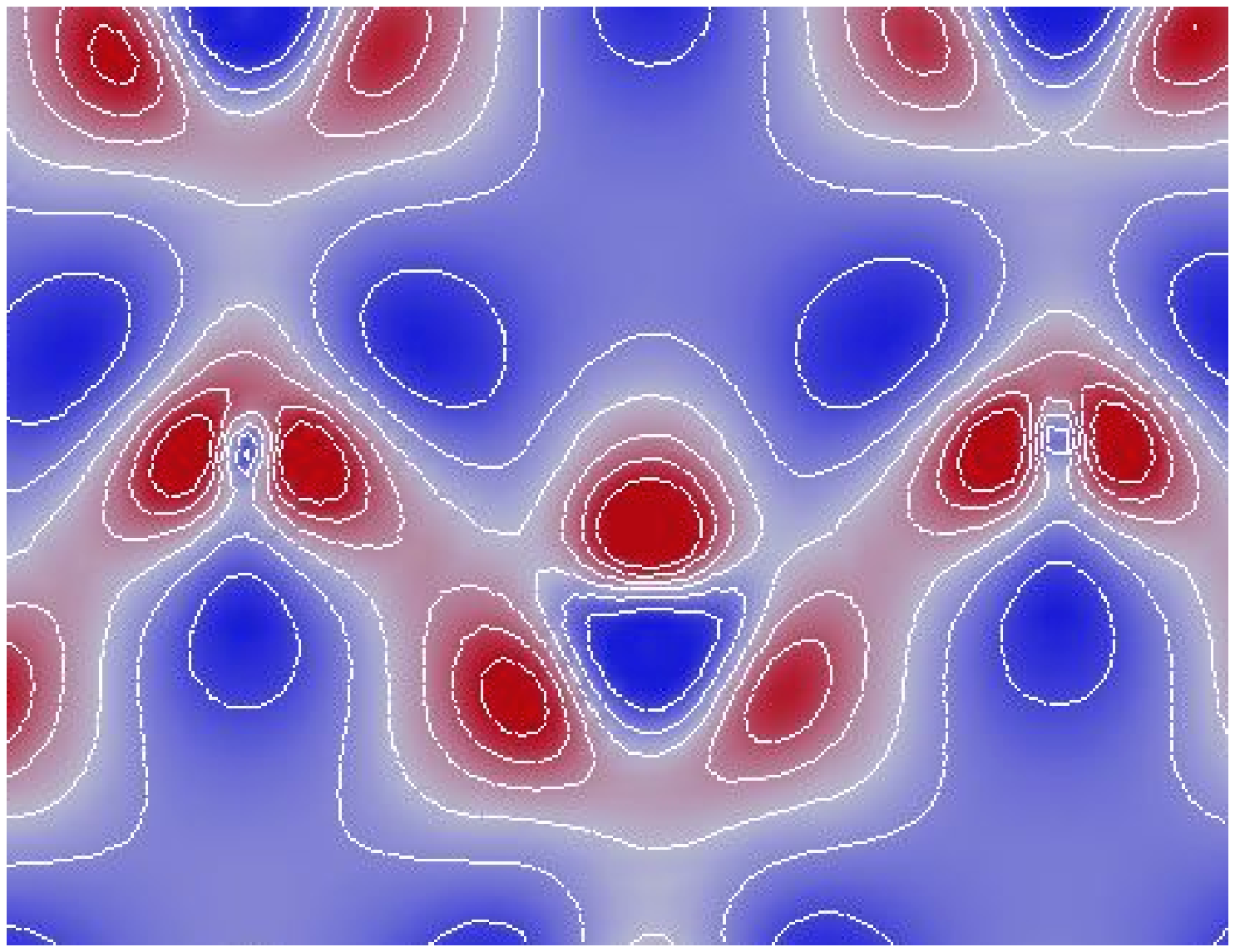}
\includegraphics[width=3.6cm,draft=false]{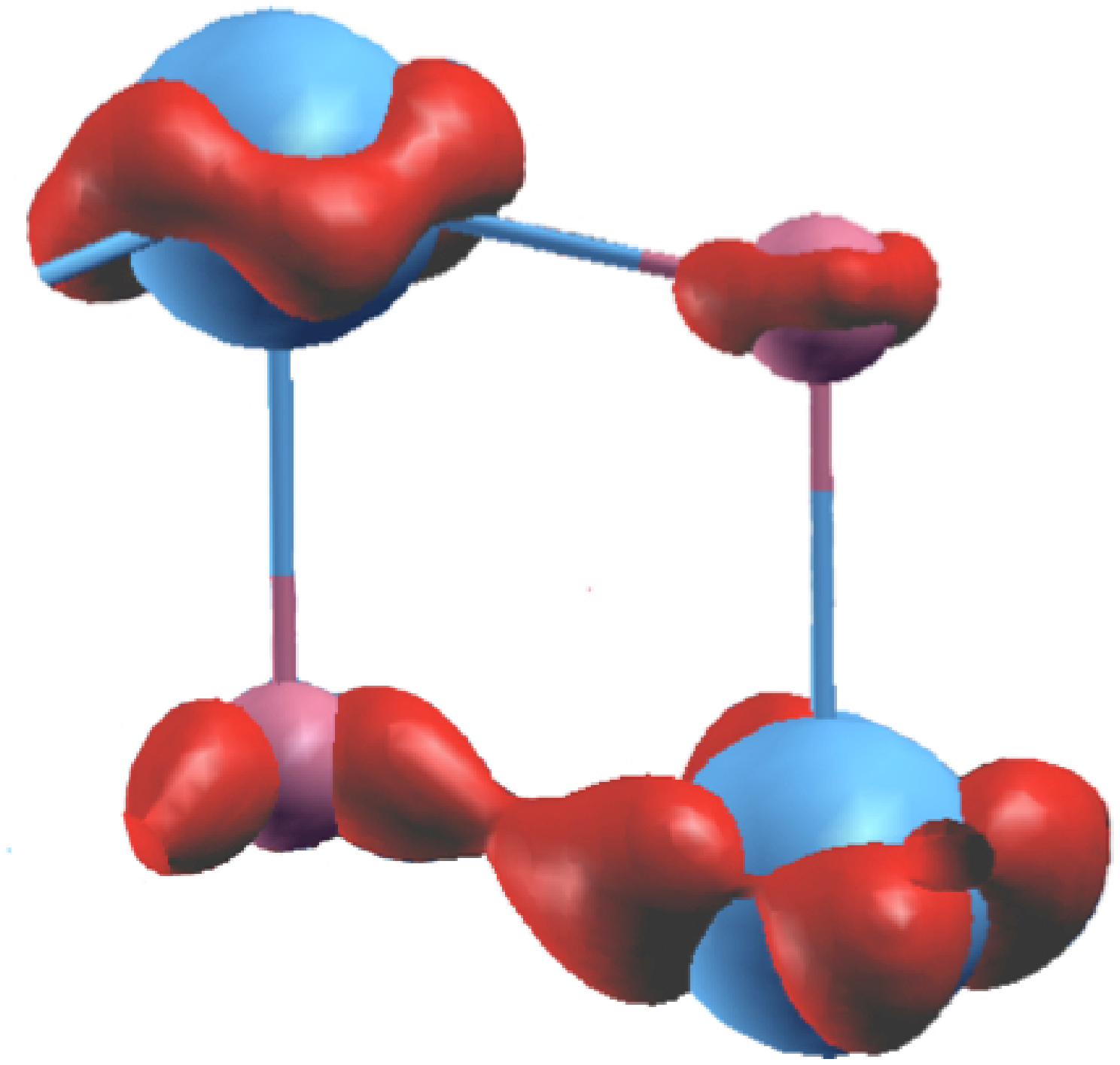}
\includegraphics[width=3.8cm,draft=false]{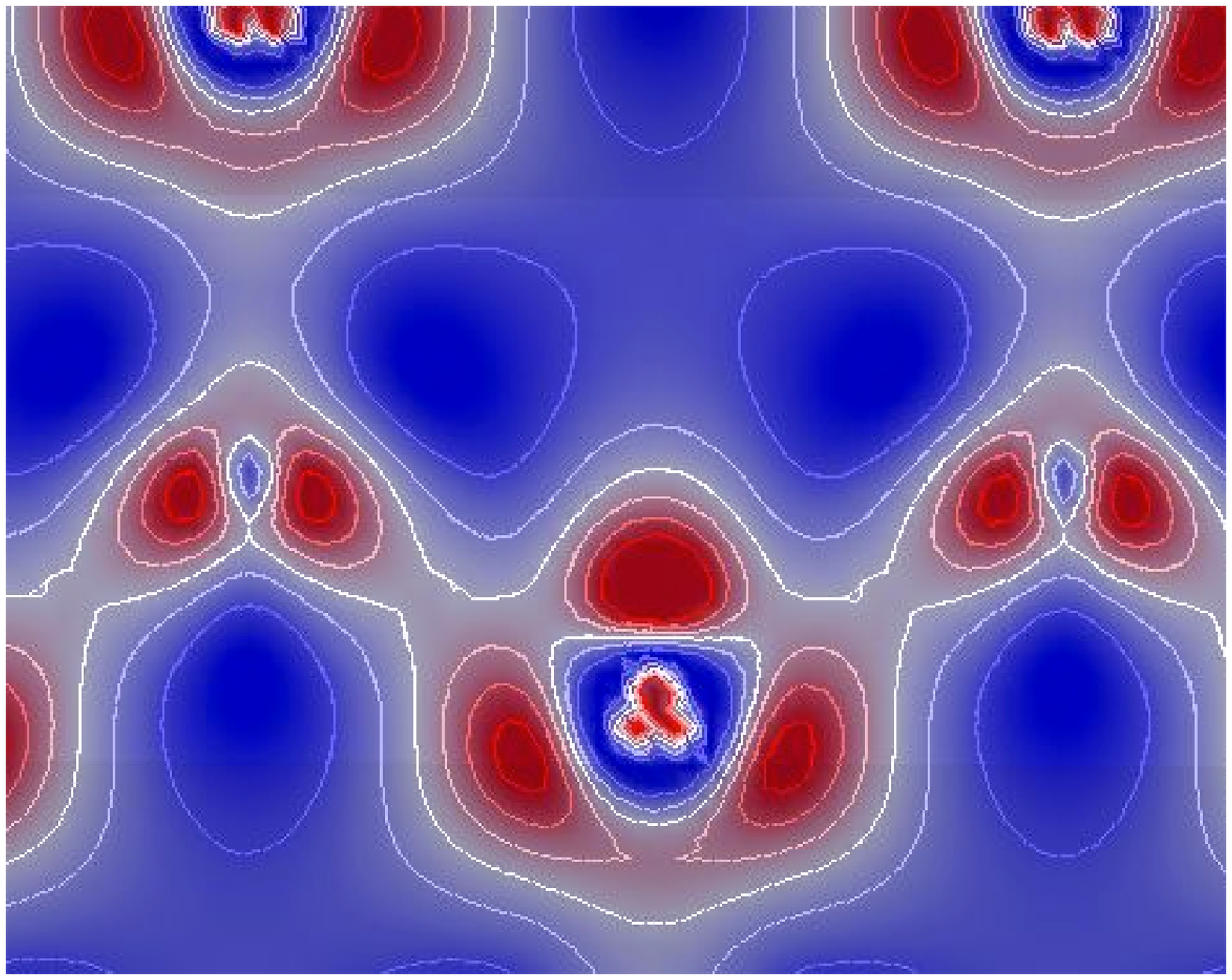}
\includegraphics[width=3.6cm,draft=false]{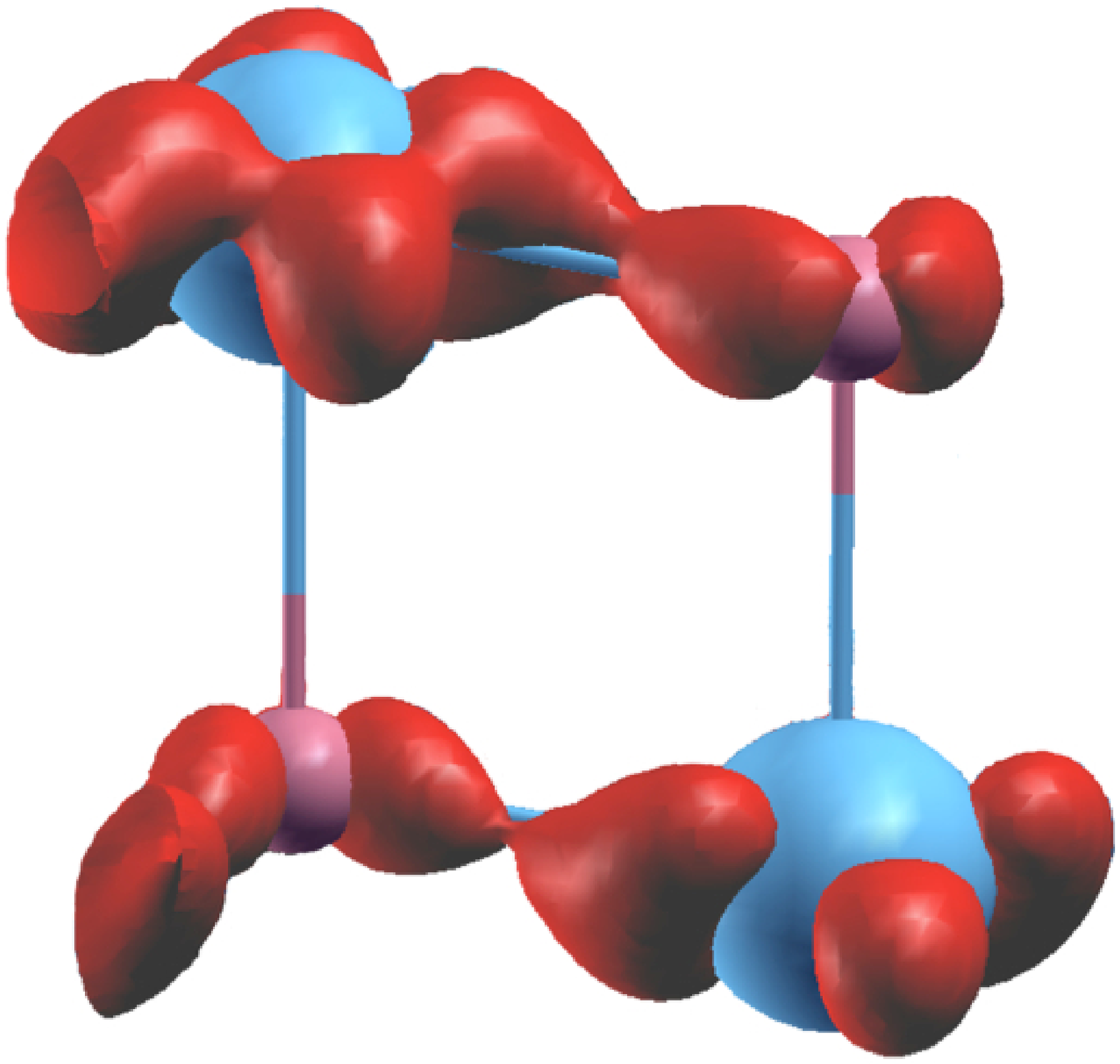}
\includegraphics[width=3.8cm,draft=false]{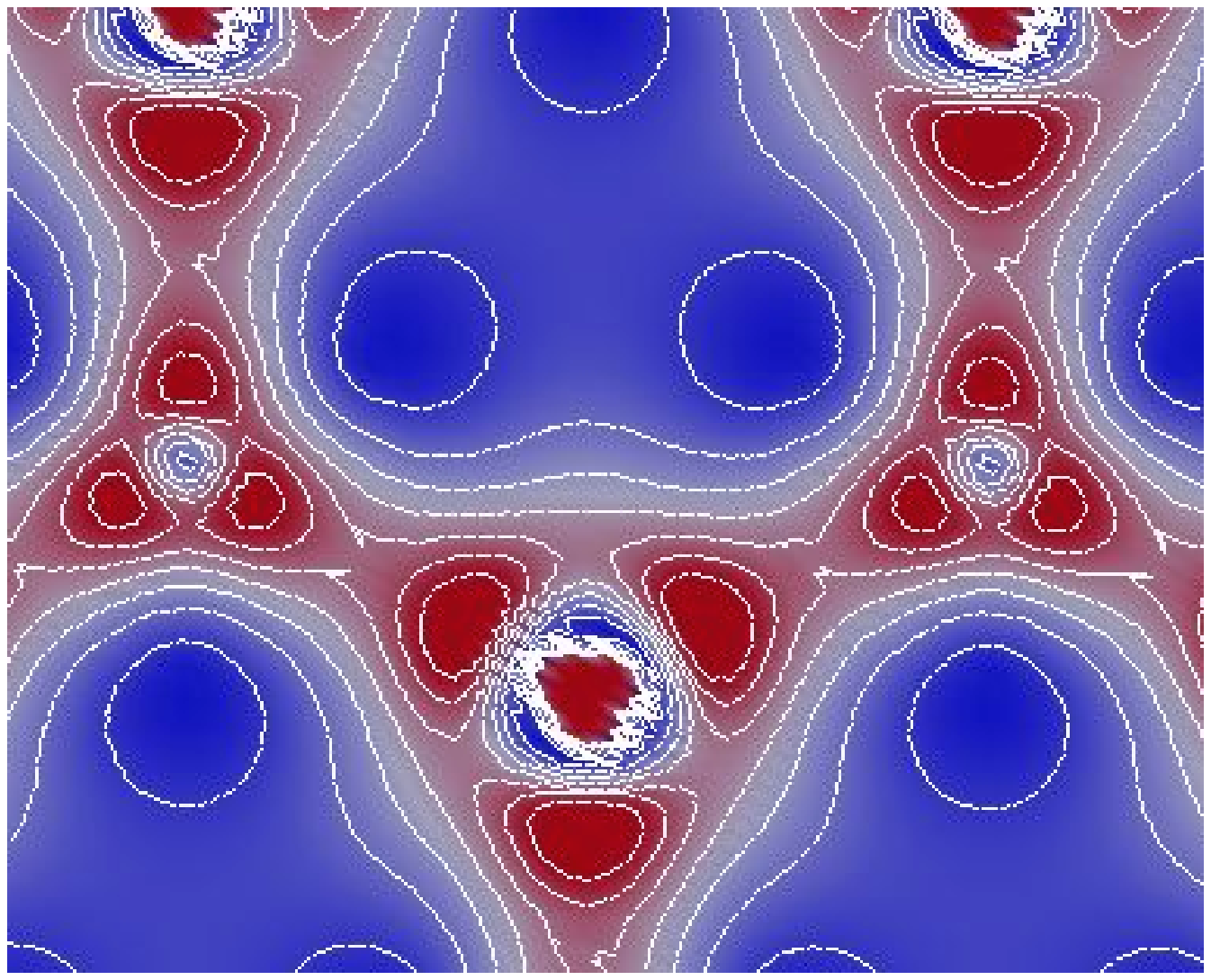}
\includegraphics[width=3.6cm,draft=false]{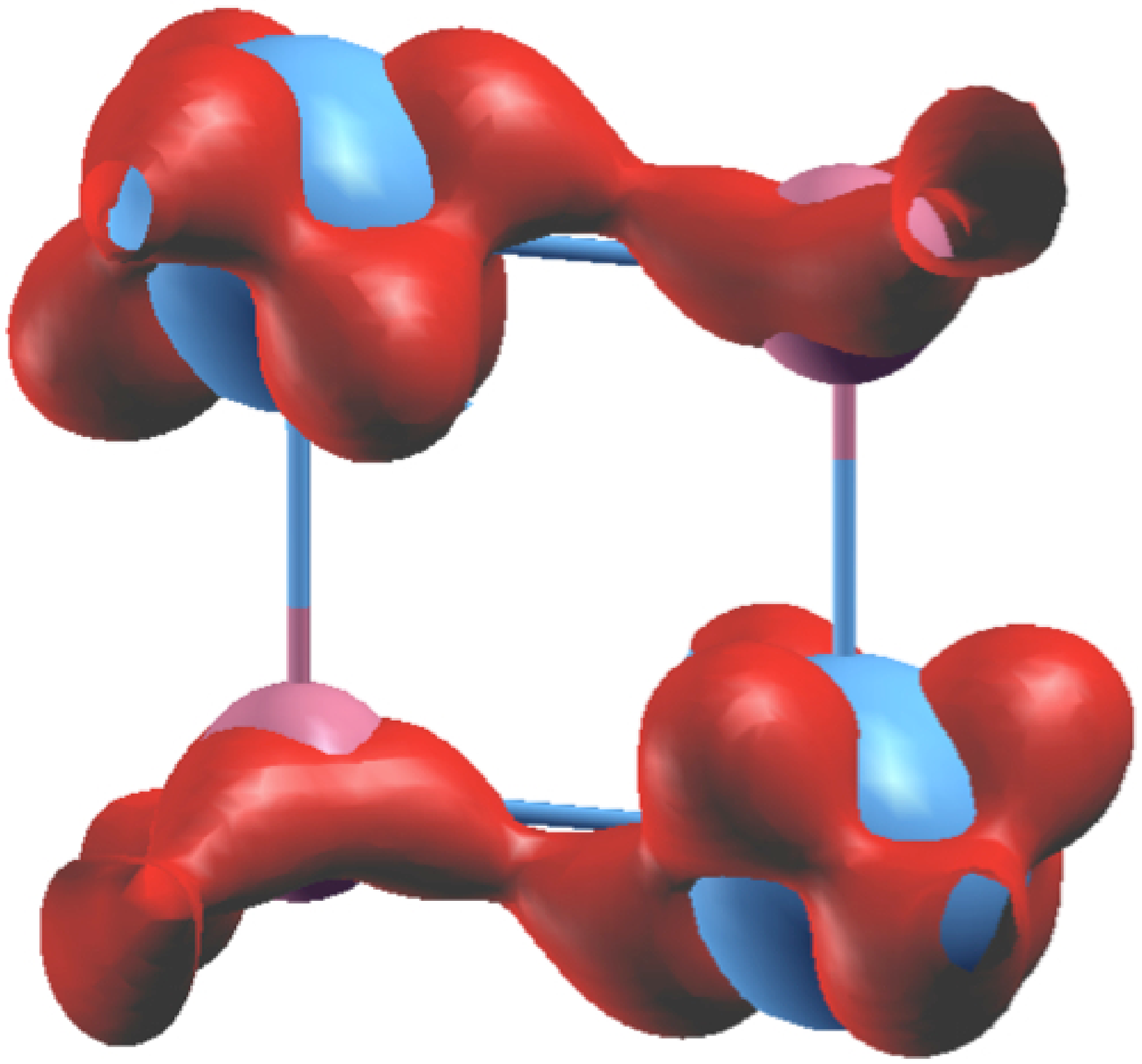}

\caption{Left panels: contour plots for Li$_x$ZrNCl for increasing values of $x= 0.16, 
0.25$ and $0.50$ from top to bottom. The plane of the figure lies between Zr and N 
layers, which are only roughly coplanar. Right panels: the corresponding three 
dimensional isocontour plot of the electron density obtained using 
Xcrysden.\cite{xcrysden} The blue and pink spheres represent Zr and N
respectively.}\label{contour}
\end{center}
\end{figure}

In Fig. \ref{contour} we show a three dimensional (3D) isocontour plot of the doped electron 
density for each non zero value of $k$ (from only the conduction bands) as well as a contour plot in a plane lying 
within one ZrN layer. The density at small doping was earlier characterized as Zr 
$d_{x^2-y^2}$, $d_{xy}$ and N $p_x, p_y$ character.\cite{weht} For $x=$ 0.16 and 0.25, the 3D plot
demonstrates that this characterization is
incomplete, because the lobes on Zr lie alternately above and below its equatorial plane.
Thus there is significant $d_{xz}, d_{yz}$ mixture due to structural asymmetries. 
The contour plots confirm that the maxima around the Zr atoms are oriented
between, rather than towards, the nearby N anions, a simple crystal field effect.

For $x$ = 0.5 new bands contribute to the density (see Fig. \ref{dos_zrncl}) 
which is reflected in a dramatic
reorientation of the high density lobes being directed towards N, with formation of a
more bonding-like structure between Zr and N.  This substantial change is not reflected
in superconducting behavior however, as T$_c$ is observed to depend hardly at all
on the doping level in the range we have considered. \cite{interlayer_coupling_2}

\section{Dielectric response of Li$_x$ZrNCl}\label{optic}

For the symmetry of Li$_x$ZrNCl the dielectric tensor is diagonal with only in-plane $xx$ and perpendicular $zz$ components. 
For simplicity, and because electronic structure differences between supercell and VCA calculations are small, we present calculations using VCA and the primitive cell to evaluate the imaginary 
part of the $q$=0 dielectric tensor within the random phase approximation (RPA).
The calculation takes the usual expression requiring the band energies and momentum 
matrix elements between Kohn-Sham wavefunctions.\cite{lindhard, lindhard_2, claudia} For x$\geq$ 0, we focus on the results obtained from VCA calculations in
which electrons are removed from Li$_{0.5}$ZrNCl. 

We have performed calculations for the polarization parallel to the ZrN planes and also for the perpendicular polarization to study anisotropy and discriminate the contribution of the ZrN planes making possible to identify the excitations in the ZrN plane and to further characterize the electronic structure of that plane for the respective doping level.

\subsection{Handling of the Intraband Part}
Without considering scattering, the intraband part is a $\delta$-function at 
$\omega$=0 with strength given by
the Drude plasma frequency $\Omega_p$ obtained from the Fermi surface average 
density of states and  velocity $\vec v_k$ averages
\begin{equation}
\Omega^2_{p,\alpha\alpha} = 4\pi e^2 N(0) v^2_{F,\alpha}.
\end{equation}
Considering that electrons are always interacting, with electrons, with phonons, and
with defects, a lifetime broadening $\gamma$ is introduced, leading to the form 
\begin{equation}
\label{im_eps}
\epsilon_{2,\alpha\alpha}^{intra}(\omega)=\frac{\gamma \Omega^2_{p,\alpha\alpha}}{\omega(\omega^2+\gamma^2)} 
\end{equation}
with Kramers-Kronig transform
\begin{equation}
\label{real_eps}
\epsilon_{1,\alpha\alpha}^{intra}(\omega)=1- \frac{\Omega^2_{p,\alpha\alpha}}{\omega^2+\gamma^2} 
\end{equation}
from the combined expression
\[\epsilon_{\alpha\alpha}^{intra}(\omega) = 1 -\frac{\Omega^2_{p,\alpha\alpha}}{\omega^2 + \gamma^2}(1 - i\frac{\gamma}{\omega})
                    = 1 - \frac{\Omega^2_{p,\alpha\alpha}}{\omega (\omega + i \gamma)}. \]
The imaginary part retains the $\omega\rightarrow$0 divergence of a metal. Given no relevant experimental information, we have used
$\gamma$ = 0.1 eV throughout, corresponding to a relaxation time $\tau$ $\approx$ 7$\times$
10$^{-15}$ s.

The calculated plasma frequencies in eV for the x= 0.21 structure (interlayer distance= 9.28 \AA) are
\begin{align}
x=0.16& ~~~\Omega_{p,xx}=2.97,&~~~ \Omega_{pl,zz}=0.70\nonumber \\
x=0.25& ~~~\Omega_{p,xx}=3.14,&~~~ \Omega_{pl,zz}=0.68\nonumber \\
x=0.50& ~~~\Omega_{p,xx}=2.80,&~~~ \Omega_{pl,zz}=0.71
\end{align}
The first two values of $\Omega_{p,xx}$ are roughly consistent with a constant
DOS with initial doping and $v_F \propto k_F \propto \sqrt{x}$ for a 2D system, as
$\sqrt{x}$ increases only from 0.4 to 0.5. For $x$=0.50, the Fermi level has moved
into additional bands and the values of the plasma frequency
cannot be estimated. The constancy of the
values of $\Omega_{p,zz}$ is consistent with the interlayer spacings which do not vary much with the Li content. They will however decrease drastically when large
organic molecules are intercalated with the Li, separating the layers considerably
and increasing T$_c$ in the process. The Li intercalation appears to enhance hopping along the $c$ axis. For the undoped compound with an interlayer distance of 9.22 \AA ~(with electrons added to simulate the desired doping level) the $zz$ values can drop by an order of magnitude or more:
\begin{align}
x=0.16& ~~~\Omega_{p,xx}=2.41,&~~~ \Omega_{pl,zz}=0.03\\
x=0.25& ~~~\Omega_{p,xx}=2.91,&~~~ \Omega_{pl,zz}=0.06
\end{align}

\subsection{Dielectric Tensor}

\begin{figure*}
\center
\includegraphics[width=7cm,draft=false]{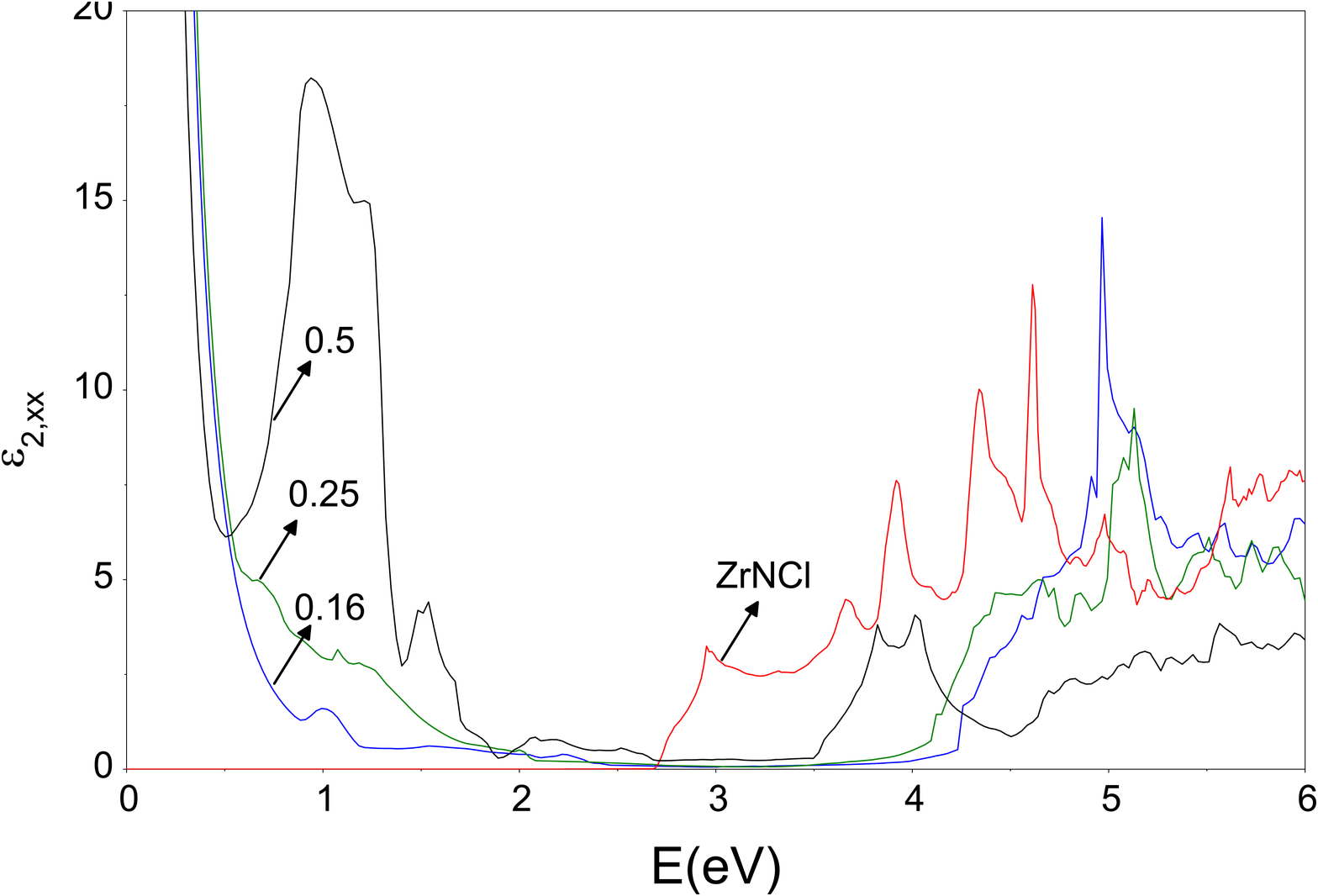}
\includegraphics[width=7cm,draft=false]{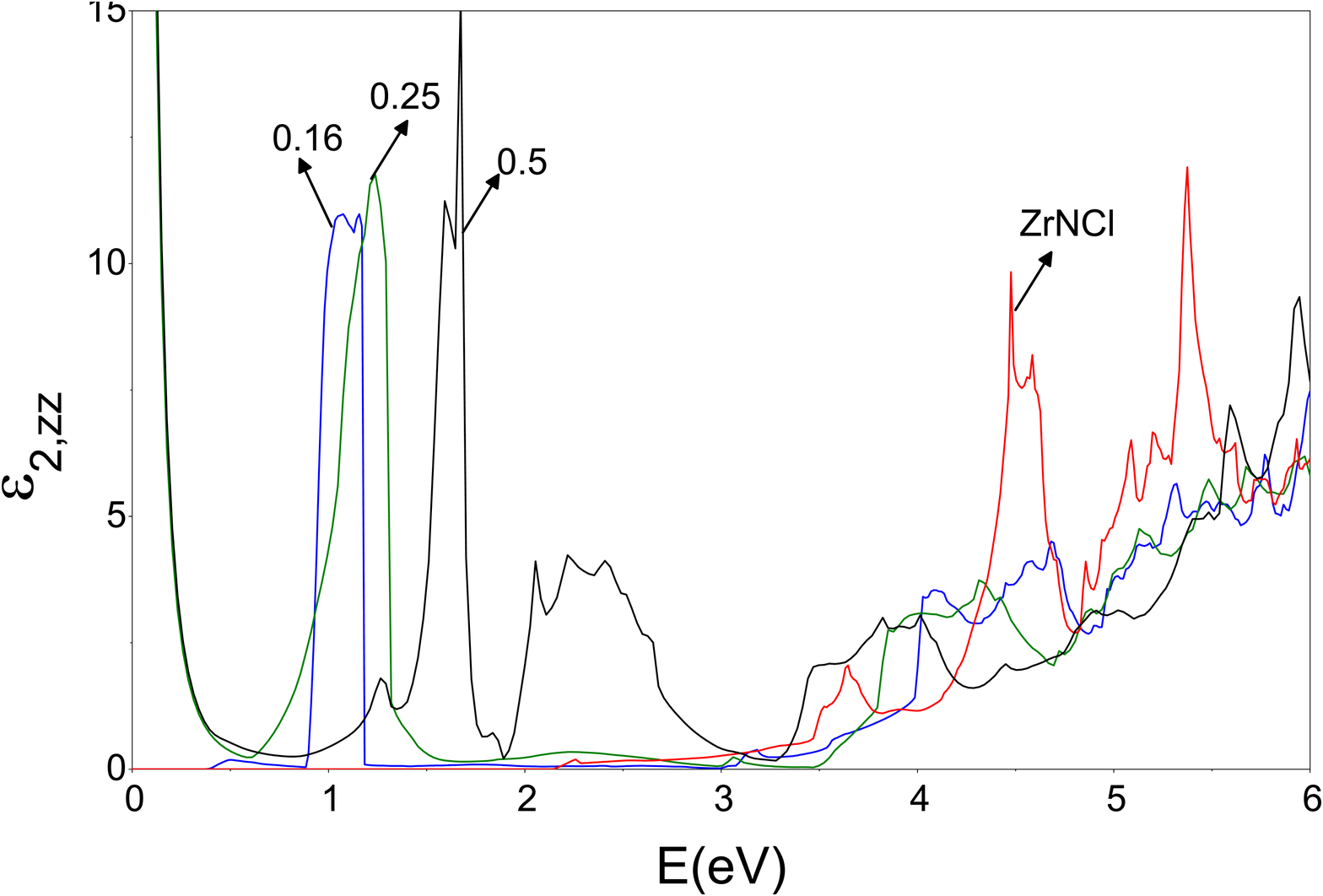}
\includegraphics[width=7cm,draft=false]{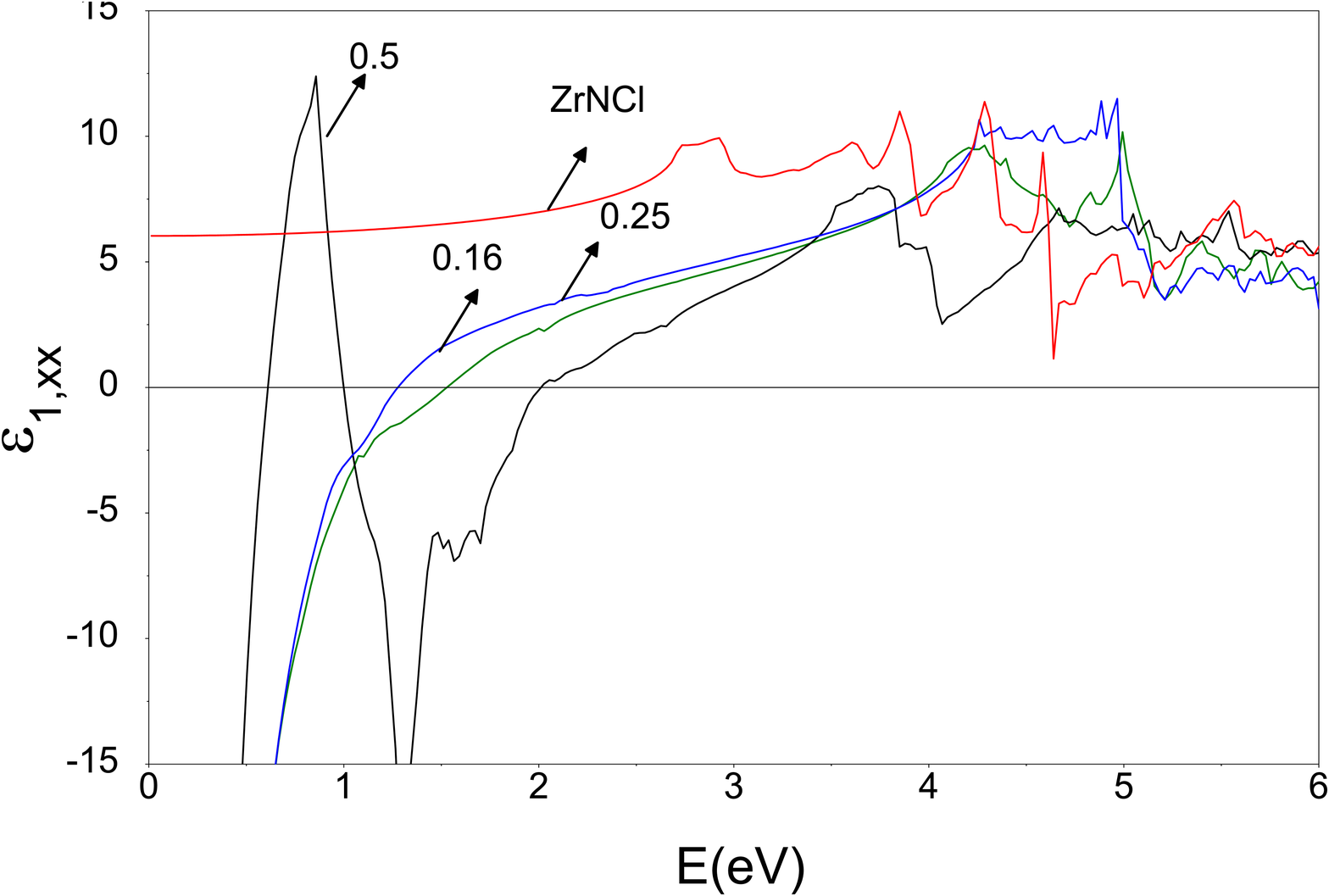}
\includegraphics[width=7cm,draft=false]{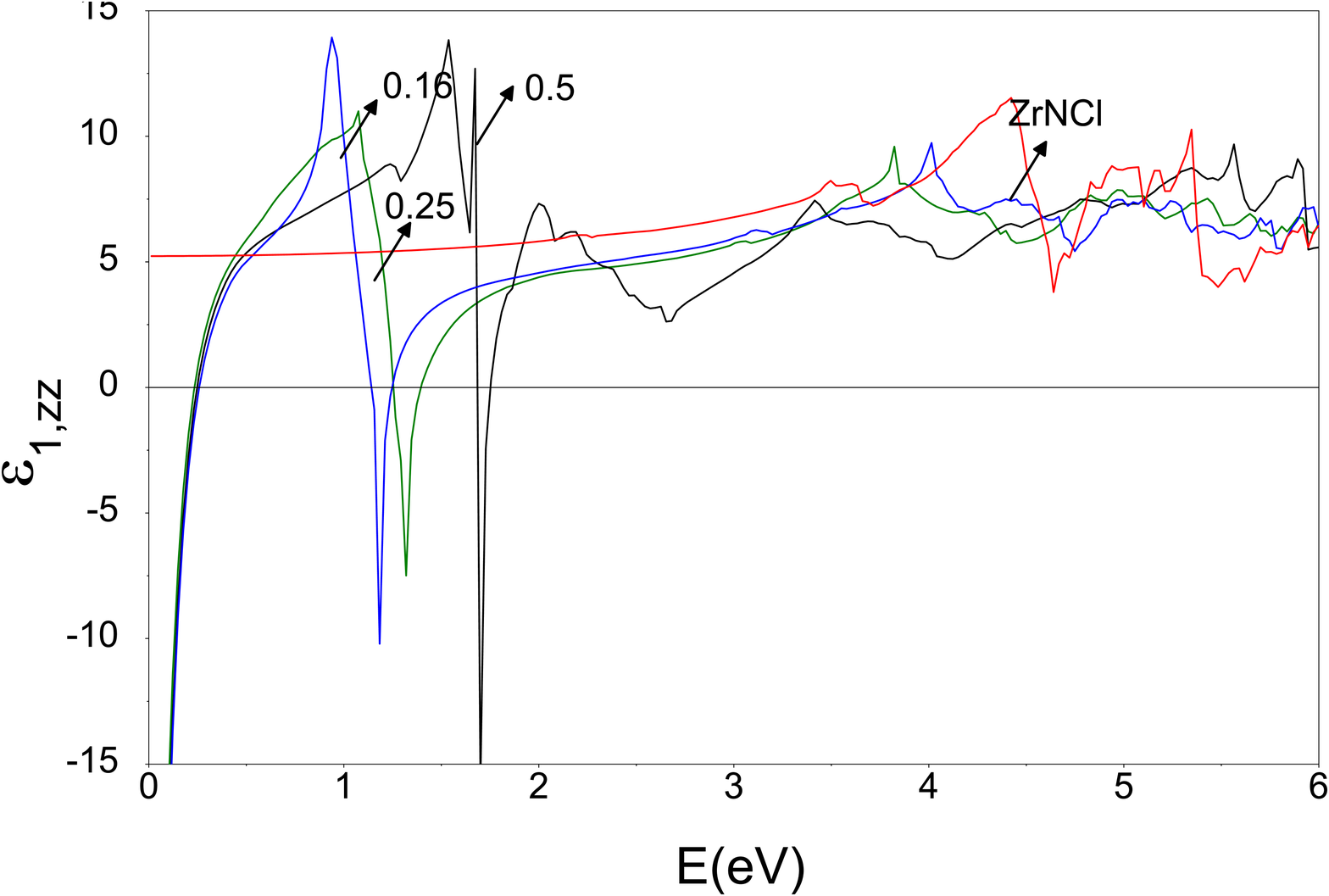}
\caption{Comparison of the real ($\epsilon_{1}(\omega)$, bottom panels) and 
imaginary ($\epsilon_{2}(\omega)$, top panels) parts of the dielectric function, 
for pristine ZrNCl and for Li$_x$ZrNCl ($x$= 0.16, 0.25 and 0.50).}\label{eps}
\end{figure*}

Fig. \ref{eps} shows the calculated dependence of the real and imaginary part of 
both components of the dielectric function for the pristine compound as well as 
for Li$_x$ZrNCl ($x$= 0.16, 0.25, and 0.50).
For undoped ZrNCl the experimental value\cite{iwasa_optic} of the static dielectric constant is $\epsilon_{1,xx}$=5. 
From the calculated electronic structure we obtain
an almost identical value. The dielectric function is featureless until the 
energy reaches the gap. Above the onset transitions from N $2p$ valence
bands to Zr $3d$ conduction bands give rise to structure.
Structures appear in $\epsilon_{1,xx}$ and $\epsilon_{2,xx}$ at energies 
in agreement with structures seen in reflectivity measurements (see more
discussion below). In $\epsilon_{1,zz}$ and $\epsilon_{2,zz}$ the response is similar with structure appearing at higher energies (above 3.5 eV). 

As mentioned above, for $x$$>$ 0, we focus on $\epsilon(\omega)$ obtained from VCA calculations in
which electrons are removed from Li$_{0.5}$ZrNCl (shown in Fig. \ref{eps}). 
In every case the optical spectrum can be divided in two regions: a lower energy 
range where peaks appear at 1-2 eV linked to interband transitions within the
conduction bands occupied by doped-in carriers, and a higher energy range from 3 eV onwards that includes
valence-conduction transitions. 

{\it In-plane polarization.} Superconductivity in this system is clearly associated with the 2D Zr-N bilayer alone
as discussed in the Introduction, so we discuss first the $xx$ component. The contribution from doped
carriers appears in the 0.7-1.4 eV region, where there is mild interband structure in $\epsilon_{2,xx}$
for $x$=0.16 and 0.25. For $x$=0.50, a dramatic change arises due to new interband transitions from the
in-plane $d_{x^2-y^2}, d_{xy}$ states to orbitals with more out-of-plane character. The changes in both
$\epsilon_2$ and $\epsilon_1$ are dramatic.  However, with no noticeable change in superconductivity in
this region of doping $x$, we conclude there is no impact of this change on superconductivity.
Doping does not make any significant change in the spectrum above 3 eV as shown in Fig. \ref{eps}: the changes at higher energy are more regular and understandable from the different level of band filling.

{\it Perpendicular polarization.}
The strong anisotropy -- difference between $\epsilon_{xx}$ and $\epsilon_{zz}$ -- reflects the 
strong dependence on the momentum matrix elements on the orbitals that are involved.
In $\epsilon_{2,zz}$  a peak at 1 eV for $x$=0.16 shifts towards higher energies with increasing $x$.
At $x$=0.50 additional strong weight appears in the 2-3 eV range.  There
is an accompanying change in the structures observed in $\epsilon_{1,zz}$. These alter
somewhat the crossings $\epsilon_{1,zz} = 0$ and affect the loss function, which we return
to below.
The region above 3 eV is altered from its behavior at $x$=0 but without any evident importance.

\subsection{Reflectivity spectrum}

Iwasa \textit{et al.} reported optical reflectivity measurements on 
Li$_x$ZrNCl,\cite{iwasa_optic} for $x$ = 0 and 0.37 with the polarization parallel to the ZrN planes.  
We have evaluated the reflectivity from the surface boundary condition expression and compared it with the available experimental results.
\begin{equation}
\label{ref}
R_{\alpha\alpha}=\left| {\frac{\sqrt{\epsilon_{\alpha\alpha}}-1}
                              {\sqrt{\epsilon_{\alpha\alpha}}+1}} \right| ^2
\end{equation}

In experiments, for the insulating compound, the reflectivity decreases slowly with decreasing photon energy in the energy region 
0.1-3 eV reflecting its band insulator nature. Prominent peaks appear above 3 eV due to 
N $2p$ -- Zr $4d$ interband transitions 
with characteristic structures at 3.7, 5.0 and 5.7 eV. In the infrared region below 
0.1 eV three peak structures due to the optical phonon modes appear. According to lattice dynamics 
calculations\cite{lattice_dynamics} four IR-active phonon modes have been proposed (20, 33, and 65 meV) -- two A 
modes (displacements along the c-axis) and two E modes (along the ab plane). 

\begin{figure}
\center
\includegraphics[width=7cm,draft=false]{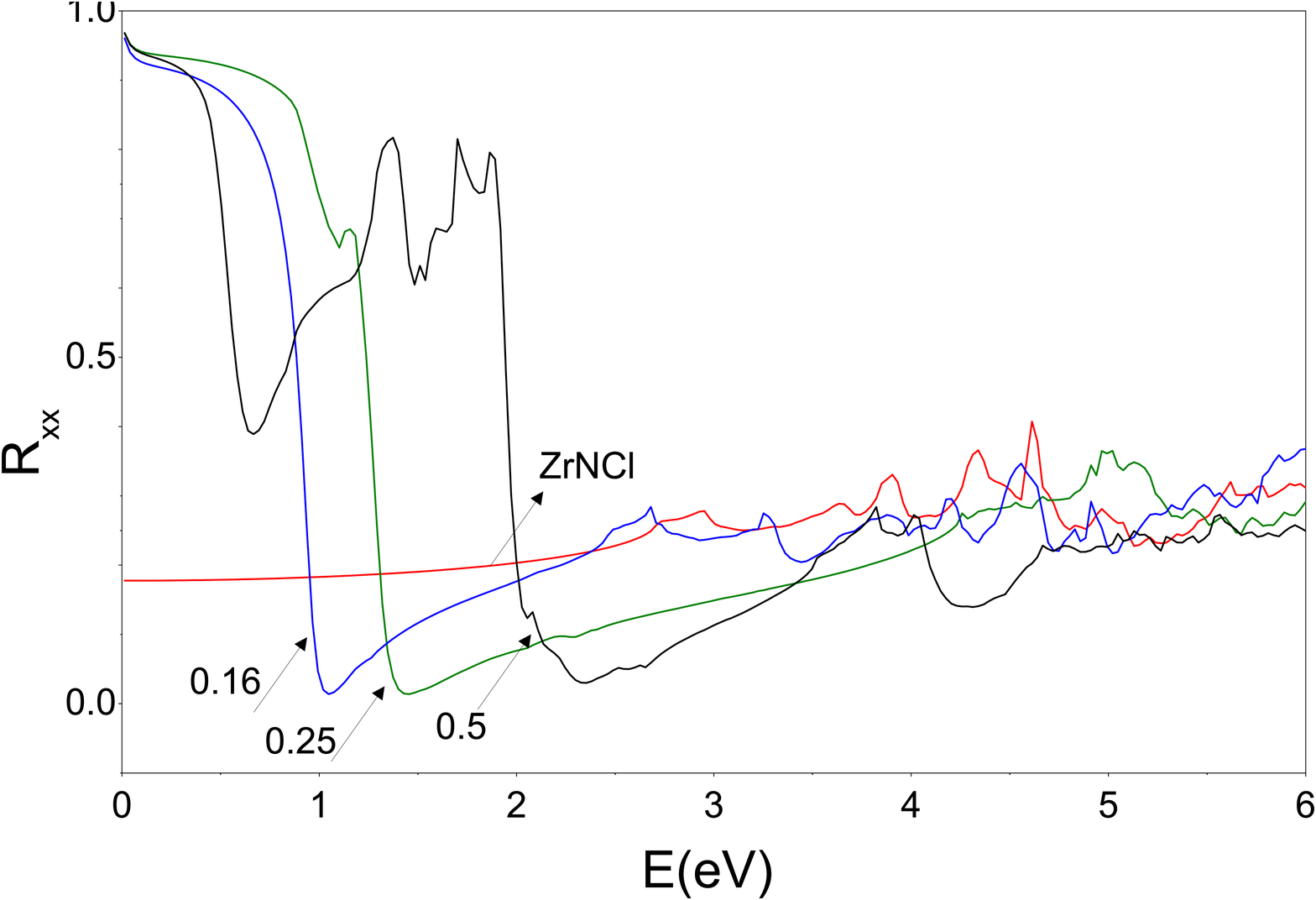}
\includegraphics[width=7cm,draft=false]{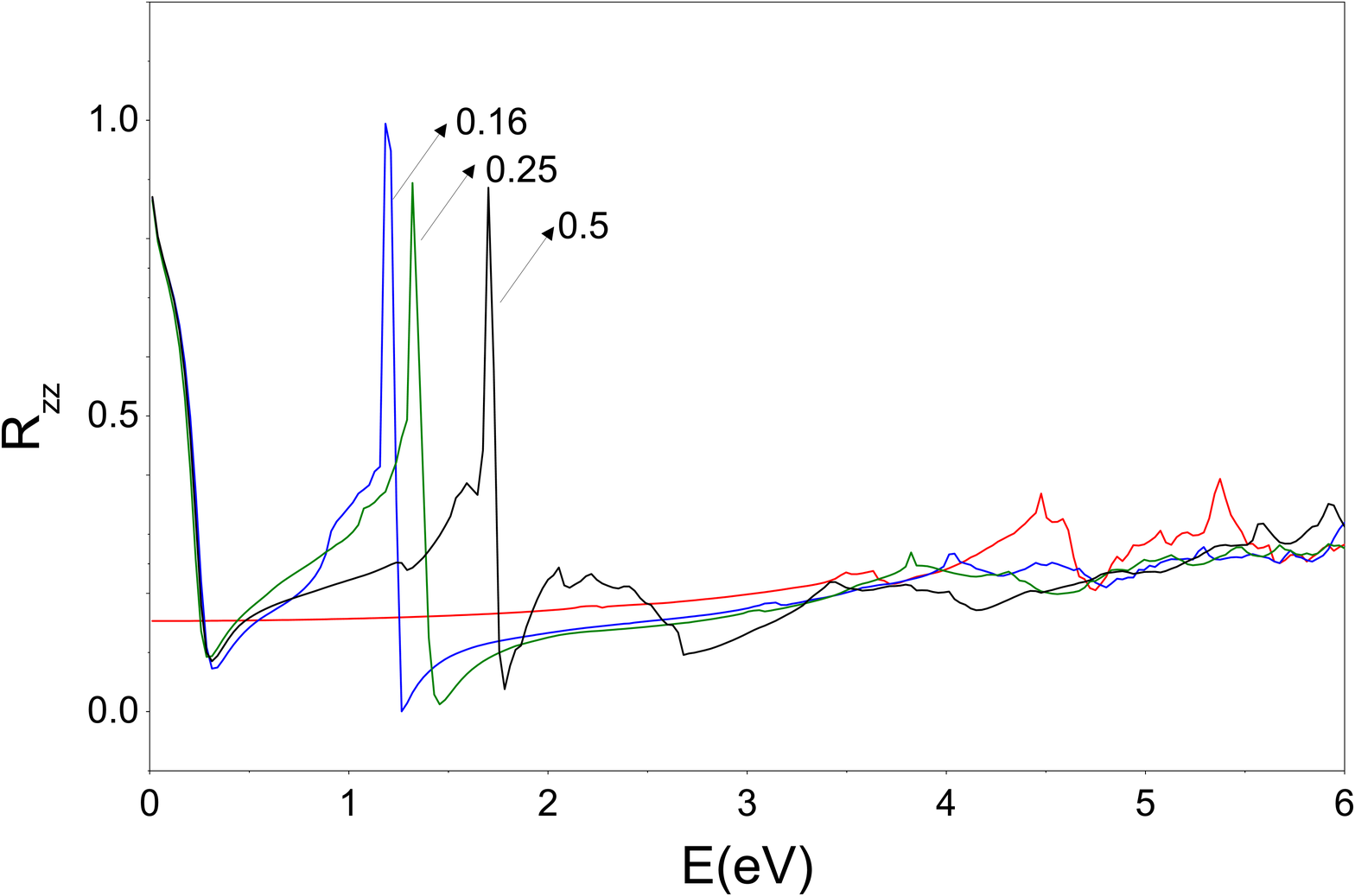}
\caption{Doping evolution of the in plane and out of plane components of the optical reflectivity for Li$_x$ZrNCl ($x= 0.16, 0.25$ and $0.50$). The curve for the pristine compound is also shown.}\label{reflectivity}
\end{figure}

Fig.\ref{reflectivity} shows the optical reflectivity spectra (for both in plane and perpendicular
component) of Li$_x$ZrNCl calculated for our chosen doping level. 

The behavior for $xx$ polarization is in agreement with the experiment reported by 
Iwasa \textit{et al.} for the undoped and doped compounds.\cite{iwasa_optic}
For the insulating compound, the reflectivity shows the similar behavior for both 
polarizations, with interband transitions appearing in $R_{zz}$ at higher energy
($\sim$4 eV) than in $R_{xx}$ ($\sim$3 eV) due to matrix element effects, consistent with the description of $\epsilon$($\omega$).

For the doped compounds the Drude-like plasma edge appears upon doping (Fig.~\ref{reflectivity})
as expected for a metal. Metallic reflection already appears below 1 eV for x= 0.16. 
As the Li doping level is increased, the plasma edge in the reflectivity spectrum shifts 
towards higher energies for the $xx$ component. The reflectivity edge is sharpened for lower doping levels and the reflectivity below the edge rapidly increases. The behavior is again different at the higher doping level studied due to the above mentioned additional interband transitions from the in-plane to the out of plane orbitals.

For $R_{zz}$ another plasma edge appears at around 0.2 eV being independent of doping level (in this range).
We discuss in the following subsection that this invariance is because this edge is
purely intraband-derived (whereas that in $xx$ polarization is affected by interband
transitions) and the corresponding Drude plasma energies do not vary with doping level.
$R_{zz}$ contains similar structure in the 1-2 eV range but it is much sharper and also shifted towards higher energies as the doping level is increased.



The behavior found for Li$_x$ZrNCl can be contrasted with what is observed in systems such as cuprates or Fe-pnictides where also the accumulated data on the normal-state properties seem anomalous in many aspects suggesting an unconventional metallic state is developed. The proximity to a magnetic ground state Fe-pnictides and cuprates makes difficult the direct comparison with non-magnetic $A_x{\cal T}$NCl superconductors in any case.

For undoped insulating La$_2$CuO$_4$\cite{uchida, timusk}, with a charge-transfer energy gap of about 2 eV, the spectrum at low energies is very anisotropic and dominated by excitations in the CuO$_2$ plane with an edge in the $xx$ spectrum appearing at 1 eV indicating the metallic state along CuO${2}$ planes. In the $zz$ reflectivity no structure corresponding to the 1 eV edge is found. In Li$_x$ZrNCl this is not the case as we have seen in the previous description. When substituting La by Sr (La$_{2-x}$Sr$_x$CuO$_4$), the spectrum for the polarization perpendicular to the planes does not change substantially in contrast Li$_x$ZrNCl: it is still featureless (typical of an insulator) in the low energy region with the optical phonons dominating
even for superconducting compositions. However, drastic changes happen in the spectrum with polarization parallel to the planes where an edge in the reflectivity appears. It is not of the usual Drude type but composed of two contributions: a Drude-like narrow one peaked at $\omega$=0 and a broad continuum centered in the mid-infrared region. The plasma edge in the 
reflectivity spectrum stays at almost the same position owing to the formation of an 
in-gap state in the charge transfer gap with doping and is not shifted as the doping level varies as happens in Li$_x$ZrNCl.\cite{uchida, kim, timusk} 

Infrared reflectivity measurements on several 122 Fe-pnictides \cite{wu} also revealed complex features in the $xx$ component of the reflectivity with two electronic subsystems existing. The one gapped due to the spin-density-wave transition in the parent materials such as Eu(Fe)$_2$As$_2$ is responsible for superconductivity in the doped compounds such as Ba(Fe$_{0.92}$Co$_{0.08}$)$_2$As$_2$, Ba(Fe$_{0.95}$Ni$_{0.08}$)$_2$As$_2$. The second subsystem gives rise to incoherent background, present in all 122 compounds, which is basically temperature independent but affected by the superconducting transition.

\subsection{Energy loss spectrum}
\begin{figure}
\center
\includegraphics[width=7cm,draft=false]{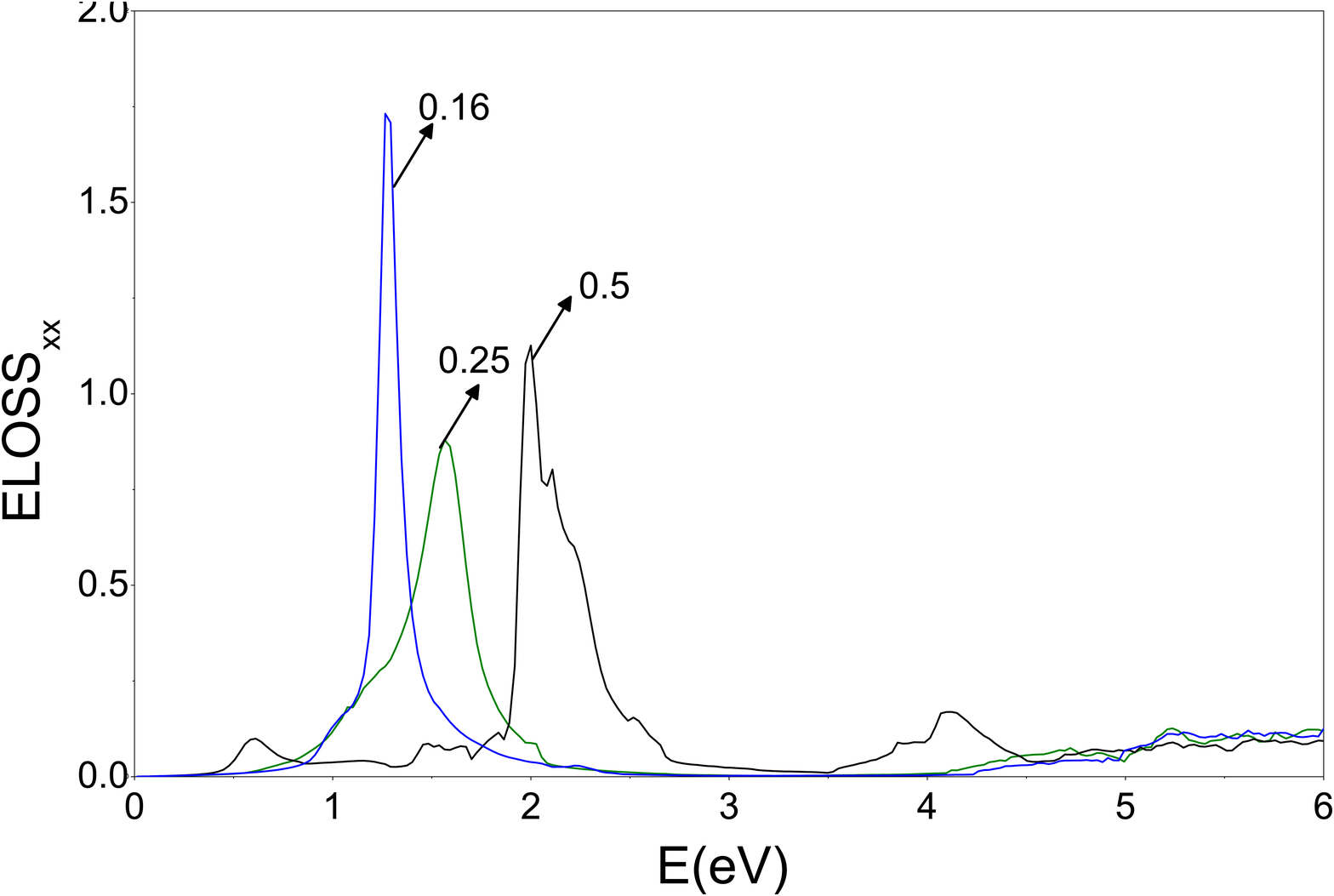}
\includegraphics[width=7cm,draft=false]{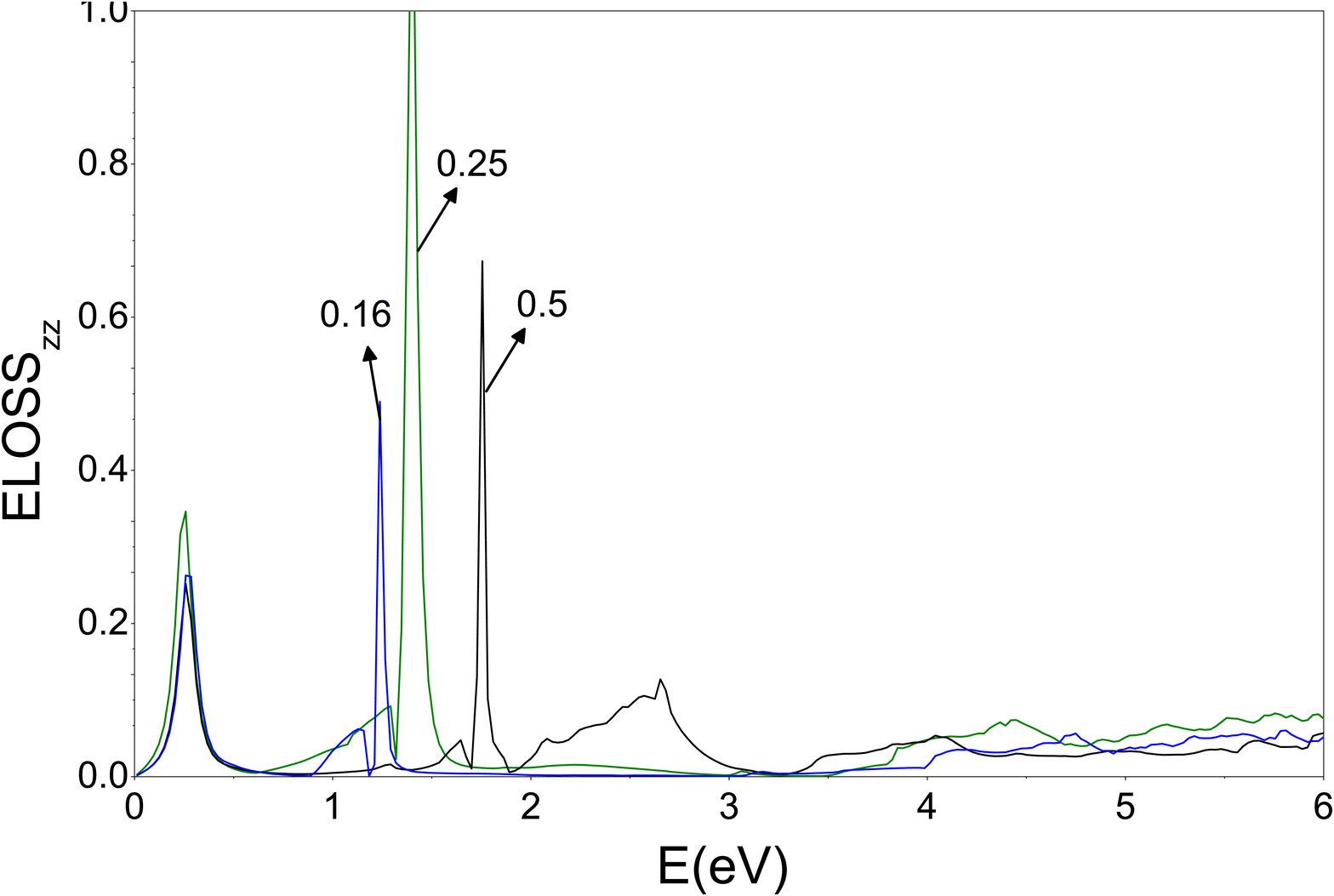}
\caption{Doping evolution of the in plane (above) and out of plane (below) components of the energy loss function of Li$_x$ZrNCl ($x= 0.16, 0.25$ and $0.50$). Note the decrease in intensity (smaller and narrower peaks) of the plasmon for out of plane polarization.}\label{eps_0_25}
\end{figure}

The energy loss function 
\begin{equation}
L_{\alpha\alpha}(\omega)=-\rm{Im} \epsilon^{-1}_{\alpha\alpha}(\omega)=
\frac{\epsilon_{2, \alpha\alpha}(\omega)}{\epsilon_{2,\alpha\alpha}^2(\omega)+\epsilon_{1, \alpha\alpha}^2(\omega)}
\end{equation}
is a characteristic response function that can be measured directly, and provides the
spectrum energy loss processes. A primary interest is in the collective 
plasmon peaks and how much
they are broadened by interband processes, and the interband losses themselves. 
Peaks occur where $\epsilon_2$ is small and $\epsilon_1$ passes through zero. Or 
returning to the reflectivity spectra, the appearance of a reflectivity edge indicates the presence of a peak in the loss function. 

In an ideal 2D interacting electron gas, the plasmon behaves as $\omega_p(q) \propto
\sqrt{q}$, hence giving no contribution at $q$=0. Doped ZrNCl however comprises a
periodic superlattice of 2D layers, and in such multilayer systems the long range
of the Coulomb interaction (across layers, independent of any electronic hopping)
pushes the $q$=0 plasmon to finite frequency\cite{jain} followed by a spectrum
at higher frequencies.  

The RPA results for $L(\omega)$ for Li$_x$ZrNCl are displayed in Fig.~\ref{eps_0_25}.
The in-plane ($xx$) spectrum is dominated by rather well defined 
plasmon peaks centered at 1.2, 1.5,
and 2.0 eV for $x$=0.16, 0.25, 0.50 respectively. The $x$=0.50 spectrum is again not quite
as simple as those for smaller $x$, containing a small peak at 0.7 eV and continuing
energy loss out to the main peak, and a further structure around 4 eV. 

The perpendicular ($zz$) loss function is different, partly reflecting the large impact
of matrix elements that depend strongly on the directionality of
the orbitals that are involved in the transition, and also
representing different physics. 
The main plasmon peaks lie at 1.2, 1.5, and 1.8 eV respectively, all lying
at roughly the same energy as for $xx$ polarization. They are however much narrower, with
full widths at half maximum of 0.1 eV or less. In addition, for each level of doping
there is a low energy loss peak around 0.2 eV with larger width. 
This peak arises from the plasma edge linked to intraband transitions appearing at the same energy in $R_{zz}$. The highest doping level $x$=0.5 shows additional interband losses in the 2-3 eV region arising from occupation of the narrower 2nd and 3rd Zr $4d$ bands.

\section{Discussion}
We have studied the $q$=0 dielectric response of insulating and doped ZrNCl in connection
with its impressive superconductivity of uncertain origin.  Screening in a layered electron
gas as a possible origin of pairing has considerable history. 
The work by Bill {\it et al.}
was discussed in the Introduction, and 
Pashitskii and Pentegov\cite{pashitskii} have
elaborated on some aspects of the plasmon mechanism of pairing. 
For the layered homogeneous electron gas, the plasmon at $\vec q\rightarrow$0 lies
at
\begin{eqnarray}
\omega_p^2&=&4\pi e^2 \frac{n/c}{m^* \epsilon_{\infty}}\\ \nonumber
\omega_p  &\approx& 2.7 \sqrt{x}~ eV
\end{eqnarray}
for parameters appropriate to ZrNCl: $m^*$=0.6, $\epsilon_{\infty}$=5, $c$=10\AA. 
Here $n$ is the 2D density of carriers per unit area.
The values are 1.1 eV, 1.4 eV, and 1.9 eV for $x$=0.16, 0.25, 0.50 respectively,
which are very similar to our computed values 1.2 eV, 1.5 eV, and 2.0 eV that can be seen in
Fig.~\ref{eps_0_25}. Note
that if the cell spacing $c$ is doubled $\omega_p$ decreases by 2$^{1/2}$; such an
increase is observed to result in an increase T$_c$ by 30\%. At this spacing and at the
critical concentration $x_{cr}$=0.06, $\omega_p \sim$ 0.5 eV.

Formulation of a realistic superconducting gap equation and hence T$_c$ is a daunting
task. Atwal and Ashcroft built a model appropriate for polarization waves of semicore
electrons and estimated that T$_c$ of a few tens of kelvins could result.\cite{atwal} 
The model of Bill \textit {et al.}\cite{bill} was adapted to doped HfNCl, concluding
that T$_c$ = 25K could readily be obtained with realistic parameters. With their
version of approximating the plasmon pairing kernel, Pashitskii and Pentegov\cite{pashitskii}
found a very strong dependence on carrier concentration (unlike observations in the
metal nitridochloride system) and could reach T$_c \sim$ 150K. 

Without building specific models and in the context of increasing T$_c$ in several
classes of one-, two-, and three-layered cuprates, Leggett focused on the Coulomb interaction
energy, first within an isolated two-dimensional electron gas, and then on the interaction
between such layers. He argued\cite{leggett} that the gain in Coulomb interaction energy
upon entering the superconducting state is greater for several layers versus a single
layer. His approach also predicted a majority of the effect to be from small-$q$ screening
(long distance interaction). 

What remains in the dynamics of these doped metal nitridochlorides (beyond the full
$q$-dependence of $\epsilon(q,\omega)$) is the background of vibrating, highly charged
ions in materials like lightly doped ZrNCl. This dynamics is implicated in the very bad
metal behavior of resistivity discussed in the Introduction. Born effective charges have been reported for ZrNCl,\cite{bohnen} with the Zr$^{4+}$ having a BEC of +2.7 in-plane, N$^{3-}$ of -2.0 and Cl$^{1-}$ of -0.7.  A significant anisotropy is found (the out of plane BEC are 1.1, -0.7 and -0.5 for Zr$^{4+}$, N$^{3-}$, and Cl${1-}$, respectively) due to the layered structure of the system. For a similar layered ionic insulator
BaHfN$_2$, the Hf$^{4+}$ ion has a BEC of +4.5 in-plane, with N$^{3-}$ values up to
-4.5 and Ba$^{2+}$ a BEC around +3.\cite{kaur} 

There will be strong Coulomb coupling between these vibrating ions and the low density
two dimensional electron gas, possibly giving rise to polaronic behavior near or below the critical doping level of 0.06. There is a large
literature on modeling polarons and on bipolaronic mechanisms of pairing. However,
specific treatments are needed: there are many cases of doped layered ionic insulators,
but only this one class of excellent superconductors with T$_c \sim$ 15-26K.  

\section{summary}

We have studied the $q$=0 dielectric response of insulating and Li-doped ZrNCl in connection
with its impressive superconductivity of unknown origin. We have revisited the electronic structure of Li$_x$ZrNCl, establishing that the differences between rigid band
modeling and virtual crystal treatment are small, and comparing the results also with
actual lithium doping using supercells. We have analyzed the dependence of the dielectric response with frequency and doping level, reproducing the experimental static dielectric constant $\epsilon_{\infty}$=5 extremely well. In the reflectivity spectra the appearance of a Drude plasma edge demonstrates the transition from a band insulator to a metal upon Li-intercalation. In the energy loss function, the main
plasmon peaks appear where the electron gas model suggests they should, in the range 1.2-2.0 eV for
$x$ varying from 0.16 to 0.50. 
The variations upon changing the doping level found in our calculations of
the reflectivity and energy loss function are not correlated with the observed experimental (non)variation of T$_c(x)$, providing useful data in the search for the pairing mechanism in this class of superconductors. Specifically, pairing based on straightforward electronic overscreening is not supported by our results.

\section{acknowledgments}

The authors have benefited from discussions with F. Gygi, D. J. Scalapino, 
M. L. Cohen, P. C. Canfield, and M. Calandra. This project was supported by the NSF Grant DMR-1207622.

\end{document}